\documentclass[aps,amsmath,twocolumn,amssymb,floatfix,showpacs,superscriptaddress,footinbibliography]{revtex4-1}
\usepackage{amsmath}
\usepackage{mathtools}
\usepackage{braket}
\usepackage{bm}
\usepackage{graphicx}
\usepackage{amssymb}
\usepackage[dvipsnames]{xcolor}
\usepackage{multirow}
\usepackage{xfrac}
\usepackage{textcomp}
\usepackage{float}
\usepackage{enumerate}
\usepackage{subfigure}

\usepackage[colorlinks=true,linktoc=page,linkcolor=green,citecolor=magenta,urlcolor=red]{hyperref}

\newcommand\bea{\begin{eqnarray}}
	\newcommand\eea{\end{eqnarray}}
\newcommand\beq{\begin{equation}}  
	\newcommand\eeq{\end{equation}}

\usepackage[normalem]{ulem}

\begin{document}
	

\title{Correlated disorder induced anomalous transport in magnetically doped topological insulators }  
\author{Takuya Okugawa}
\affiliation{Institut f\"ur Theorie der Statistischen Physik, RWTH Aachen, 
52056 Aachen, Germany and JARA - Fundamentals of Future Information Technology.}

\author{Tanay Nag}
\email{tnag@physik.rwth-aachen.de}
\affiliation{Institut f\"ur Theorie der Statistischen Physik, RWTH Aachen, 
52056 Aachen, Germany and JARA - Fundamentals of Future Information Technology.}

\author{Dante M.\ Kennes}
\email{dante.kennes@rwth-aachen.de}
\affiliation{Institut f\"ur Theorie der Statistischen Physik, RWTH Aachen, 
52056 Aachen, Germany and JARA - Fundamentals of Future Information Technology.}
\affiliation{Max Planck Institute for the Structure and Dynamics of Matter, Center for Free Electron Laser Science, 22761 Hamburg, Germany.}

\begin{abstract}
We examine the transport properties of magnetically doped topological insulator (TI) thin films subject to correlated non-magnetic disorder. For the  disorder we choose a quasi-periodic potential with a random phase. We restrict the disorder to a central region, which is coupled to two leads in a clean quantum spin Hall insulator (QSHI) state and concentrate on different orientations of the quasi-periodicity in the two-dimensional central region. In the case of a diagonally oriented or purely longitudinal quasi-periodicity we find different topological Anderson insulator (TAI) phases, with a quantum anomalous Hall insulator (QAHI), a quantum spin Chern insulator (QSCI), or a QSHI phase being realized before the Anderson insulation takes over at large disorder strength.  Quantized transport from extended bulk states is found for diagonal quasi-periodicity  in addition to the above TAI phases that are also observed for the case of uncorrelated disorder.
For a purely transverse  orientation of the  quasi-periodicity  the emerging QSHI and QSCI phases persist to arbitrarily strong disorder potential. These topological phase transitions (except to the Anderson insulator phase),  can be understood from a self consistent Born approximation.
\end{abstract}

\maketitle

\textcolor{blue}{\textit{Introduction}---}
In recent years various non-interacting quantum Hall phases such as,  the quantum anomalous Hall insulator (QAHI) \cite{Haldane88,Onoda03}, the quantum spin Hall insulator (QSHI) \cite{Kane05a,ShengDN2006,Bernevig06} as well as Weyl and Dirac semimetals  \cite{Hasan10,Liu16,Armitage2018} have been identified. The
QSHI phases have been proposed theoretically \cite{Bernevig1757,zhang2009topological} as well as observed experimentally \cite{konig2007quantum,roth2009nonlocal} in non-magnetic materials for example, HgTe and Bi$_2$Te$_3$. 
Interestingly, the QSHI turns into a QAHI, harbouring chiral edge states,
when time reversal symmetry (TRS) is broken \cite{Mong2010,otrokov2019prediction,liJH2018,tang2016dirac,liu2016quantum,he2018topological,Saha21}. This can be achieved by magnetic doping \cite{Liu08,Li13}, exchange field \cite{Qiao10,Yang11} and staggered magnetic flux \cite{luo2017time} and was recently experimentally realized  \cite{yu2010quantized,chang2013experimental,Checkelsky2014Trajectory,chang2015high,FengYang2015,Kou2014,Bestwick2015,Chang2016Obse,Yasuda2017Quantized,sharpe2019emergent,serlin2020intrinsic}.

Remarkably, the QSHI phases  are robust against weak non-magnetic disorder while  moderate
disorder can induce a topological phase, 
called a topological Anderson insulator (TAI),
even if the clean system remains a normal insulator (NI) \cite{Li2009,Groth2009,Guo10}. The disorder results in a negative mass term for 
the band inversion as captured by the self consistent Born approximation (SCBA).  Importantly, 
weak magnetic disorder is shown to stabilize the QAHI phase while the AI phase emerges for substantially 
strong disorder \cite{Namura2011,LuHZ2011,QiaoZH2016,ChenCZ2019,Haim2019,xing2018influence,WangJ2014Uni,Keser2019,wang2018direct,lee2015imaging,lachman2015vis,kou2015metal}. However, non-magnetic disorder, originating from the spatial inhomogeneities constitutes an important factor in  experiments as well \cite{wang2018direct,lee2015imaging,YuanYH2020,chen2015magnetism,Chang2016Obse,LiaoJ2015} and has attracted recent attention \cite{Yanxia11,Prodan11,Yamakage11,Yan-Yang12,Dongwei12,Okugawa20,Zhi-Qiang21}. 
The nature of the various TAI phases also depend on the types of disorder such as, site versus bond disorders \cite{Song12}.

Interestingly, the topological phase transitions (TPTs) in the presence of both  non-magnetic 
correlated disorder, caused by quasi-periodic Aubry-Andr\'e-Harper potential \cite{aubry1980analyticity} in the two dimensional (2D) plane, and 
the magnetic exchange field remain uncharted so far while the effect of on-site random disorder \cite{Okugawa20}, correlated disorder \cite{Girschik13,Fu21}, and magnetic disorder \cite{ChenCZ2019,Haim2019}  are investigated separately. Also, 
the random disorder effects on topological Penrose-type quasicrystal  systems~\cite{Dong-Hui19} and magnetic Weyl semimetal in the presence of intra and inter-orbital disorder~\cite{Rui18} have been already studied. In particular,
we answer the following
question which is experimentally relevant as well \cite{chang2013experimental,Checkelsky2014Trajectory}: how can we understand the rich interplay between magnetism and orientation  of correlated disorder in the 2D plane by examining the topological phase diagram?
The existence  of  a mobility edge
in one-dimensional (1D)  quasi-periodic systems \cite{Biddle11,Modak15,Modak20,Deng19,Hepeng19,Liu_22} further motivates us to explore its connection with the  
edge transport in 2D topological systems.

\begin{figure}[t]
{\includegraphics[width=\columnwidth,clip]{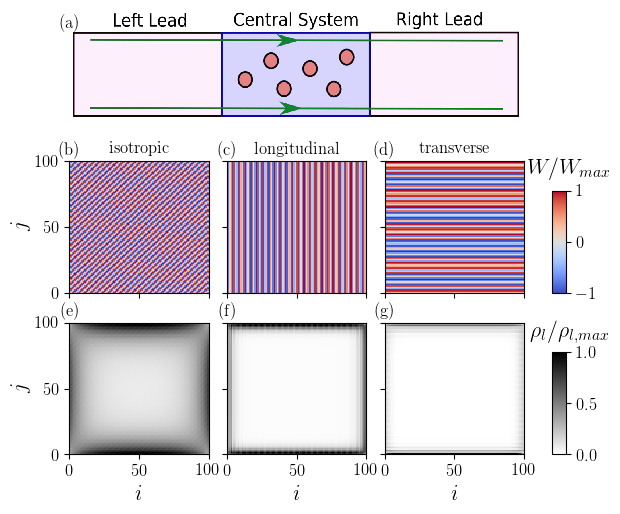}}
\caption{(Color online) The studied setup is demonstrated in (a). 
The spatial representation of quasi-periodic potential $\epsilon_r=\cos(2 \pi \eta r  )$ and disorder averaged local density of states  from the retarded Greens function of the central system are shown for
isotropic (b),  (e); longitudinal (c), (f) and  transverse (d), (g) cases. Here, $r$ represents the location of lattice sites on the 2D square lattice. We choose $gM=30$ {and $W=350$} meV  to depict edge  modes at $j=1,~100$ for the QAHI phase in (e)-(g).}
\label{system}
\end{figure}
\begin{figure*}
	\centering
	\subfigure{\includegraphics[width=\textwidth]{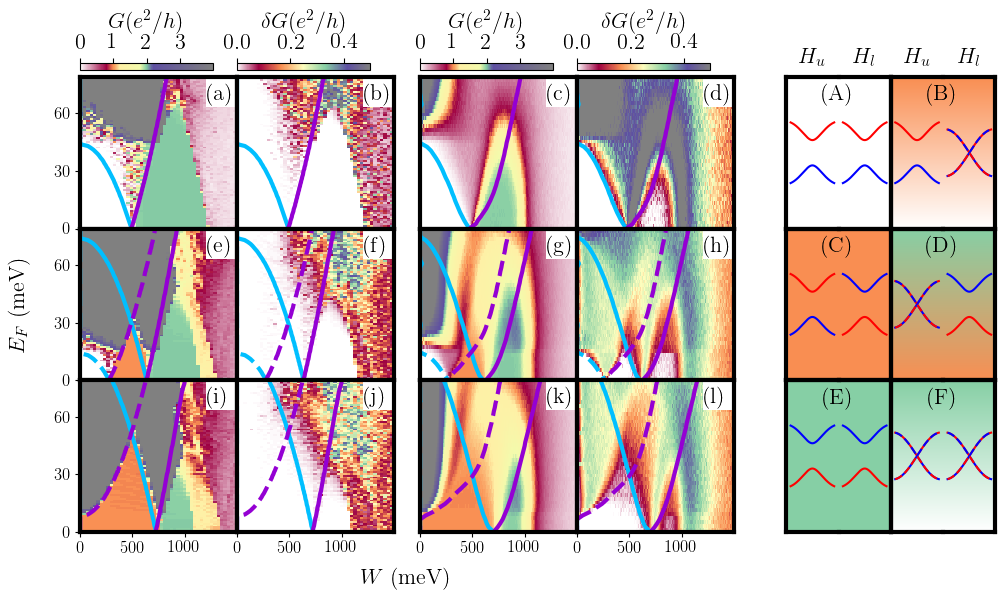}}
	\caption{(Color online)  (a), (e), and (i) [(c), (g), and (k)] depict the conductance $G$ with $gM=0$, $30$ and $52$ meV, respectively,
	for isotropic correlated disorder $\epsilon_r=W\cos(2 \pi \eta r +\phi)/2$ with $\alpha=\beta=1$ [uncorrelated random disorder $\epsilon_r \in [-W/2,W/2]$ \cite{Okugawa20}]. The corresponding standard deviation $\delta G$ are shown for isotropic correlated [uncorrelated random] disorder in 
	(b), (f), and (j) [(d), (h), and (l)].
	We consider a central system $H_{\rm CS}(m_0,M,W)$ of dimension $L_x \times L_y=400a \times 100a$ 
    and $400a \times 200a$, respectively, for zero and non-zero values of $gM$. The solid and dashed  [blue and purple] lines trace out the phase boundaries associated with $H_u (\bm k)$ and  $H_l (\bm k)$ [$|\overline{E}^{u,l}_F|=\overline{m}_{0}^{u,l}$ and $|\overline{E}^{u,l}_F|=-\overline{m}_{0}^{u,l}$], respectively, according to the SCBA.
	The cartoon pictures (A)-(F), depict the evolution of the band gap for topological (red valence and blue conduction bands in the uniform orange and green background) and trivial (blue valence and red conduction bands in the white background) phases for $H_{u,l}(\bm k)$ separately.
	}  
	\label{xy_correlated}
\end{figure*}
\begin{figure}
	\centering
	\subfigure{\includegraphics[width=\columnwidth]{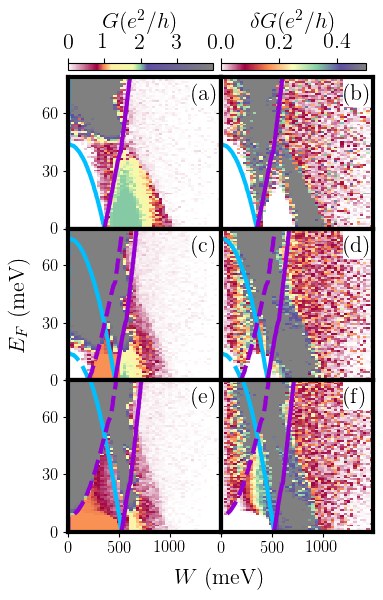}}
	\caption{(Color online) We investigate the longitudinal quasi-periodicity, $\alpha=1$ and $\beta=0$ for the same set of parameters as Fig.~\ref{xy_correlated}. The size of the QSHI (QSCI) phase for zero (non-zero) values of $gM$ reduces as compared to the Fig.~\ref{xy_correlated} (a) and (e). 
	}
	\label{x_correlated}
\end{figure}
\begin{figure}
	\centering
	\subfigure{\includegraphics[width=\columnwidth]{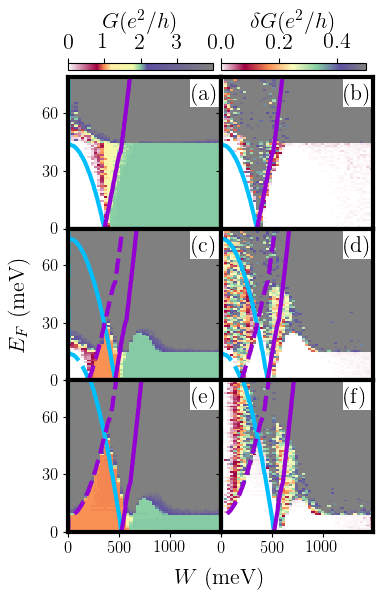}}
	\caption{(Color online) We investigate the transverse  quasi-periodicity, $\alpha=0$ and $\beta=1$ for the same set of parameters as Fig.~\ref{xy_correlated}. The QSHI (QSCI) phase remains stable even at strong disorder unlike in the other two cases shown in Fig.~\ref{xy_correlated} and \ref{x_correlated}.
	}
	\label{y_correlated}
\end{figure}

In this work, we consider Bi$_2$Te$_3$ thin film
in the presence of magnetic exchange field and correlated disorder (here, chosen as a quasi-periodic potential with random phase), coupled to two semi-infinite clean non-magnetic leads in the QSHI phase, to investigate the conductance 
through the former. The disorder, depending upon its orientation (see Fig.~\ref{system}),  can mediate a series of TPTs as observed in the  rich  phase diagrams
where the system transits through a number of phases such as, NI, QAHI, QSHI, quantum spin Chern insulator (QSCI), and AI phases.    
The diagonal quasi-periodic case, called isotropic in the following, surprisingly yields quantized conductance from the extended bulk states beside the TAI (QAHI, QSHI and QSCI) phases with quantized edge transport (see Fig.~\ref{xy_correlated}). For anisotropic longitudinal [transverse] quasi-periodicity, the QSCI phase gets remarkably suppressed [extended] when the exchange field and disorder amplitude
increase (see Fig.~\ref{x_correlated}) [see Fig.~\ref{y_correlated}].  
The TPTs in the above cases are successfully captured by the sign change of renormalized mass term computed from SCBA.

\textcolor{blue}{\textit{Model and method}---}
We start with a model of three quintuple layers of (Bi, Sb)$_2$Te$_3$ given by \cite{Bernevig1757,wang2015,SuppMater} 
\begin{equation}
 H_0(\bm k)= {\bm N}\cdot {\bm \Gamma}=\sum^3_{i=1}N_i\Gamma_i
 \label{eq:model1}
\end{equation}
where $N_1=v_F \sin(k_y a)/a$, $N_2=-v_F \sin(k_x a)/a$, $N_3=m(\bm{k})=  m_0 + 2B[2- \cos(k_x a)-\cos(k_y a)]/a^2$ and $\Gamma_1= \tau_x \sigma_0$, $\Gamma_2= \tau_y \sigma_z $, and $\Gamma_3= \tau_z \sigma_0$. We note that ${\bm \tau}$ and ${\bm \sigma}$ represent orbital and spin degrees of freedom. 
Here, $v_F$ ($a$) denotes the Fermi velocity (lattice spacing).  
A ferromagnetic order in the above topological insulator (TI) thin film can be induced by  magnetic doping with Cr or Fe atoms
\cite{chang2013experimental,chang2015high,wang2015}. Such a TRS broken TI can be modeled as  $H(\bm k)=H_0(\bm k)+  gM \tau_z \sigma_z$ where $g$ represents the Lande-g factor and $M$ is the   magnetic exchange field. The Hamiltonian thus reads in the block-diagonal form \cite{yu2010quantized}: 
\begin{equation}
{ H}(\bm k)=
\begin{pmatrix}
{H}_u(\bm k)
 & 0 \\
0 & {H}_l(\bm k)
\end{pmatrix},
\label{hamiltonian} 
\end{equation}
where the upper and lower block Hamiltonian ${H}_{u,l}(\bm k)=N_{\mp} \sigma_+ + N_{\pm} \sigma_- + m_{u,l}(\bm k) {\sigma}_z$ with $N_{\pm}=N_1 \pm i N_2$, $\sigma_{\pm}=(\sigma_x \pm i \sigma_y)/2$ and $m_{u,l}(\bm k)=N_3 \pm gM$. 


We study the effect of an on-site non-magnetic impurity potential on the central magnetic TI  (Eq.~(\ref{hamiltonian})) of dimension $L_x\times L_y$
which is coupled to semi-infinite QSHI leads (Eq.~(\ref{eq:model1})). The setup is shown in Fig.~\ref{system} (a).
We model the impurity potential by a quasi-periodic potential with random phase $\epsilon_{i,j}=W \cos{\bigl[2 \pi \eta (i \alpha + j \beta) + \phi\bigr]}/2$, where $W$ denotes the amplitude of the potential i.e., disorder strength, $\phi$ is an offset chosen from a uniform random distribution between $[0,2 \pi)$ and $\eta=(\sqrt{5}-1)/2$ is an irrational number. The disorder correlation function takes the form $C_{m,n} =\left\langle \epsilon_{i,j} \epsilon_{i+m, j+n} \right\rangle = W^2 \cos{\bigl[2\pi\eta (m\alpha +n \beta)\bigr]}/8$. Owing to the finite value of $C_{m,n}$ for quasi-periodic potential with random phase, unlike the random potential with $C_{m,n}=\delta_{n,0}\delta_{m,0}$, we refer to $\epsilon_{i,j}$ as correlated disorder.
The real space Hamiltonian for the central system and the leads are given by $H_{\rm CS}(m_0,M,W)= \sum_{r r'}[ {\mathcal H}_{rr'}(m_0,M) + \epsilon_r \delta_{r, r'}]C^{\dagger}_r C_{r'} $ and $H_{L,R}(m_0)=H_{\rm CS}(-|m_0|,M=0,W=0)$, respectively, where ${\mathcal H}_{rr'}(m_0,M)$ is obtained from a Fourier transformation of ${ H}(\bm k)$ in Eq.~(\ref{hamiltonian}) where $r$ represents the location of lattice sites on the 2D square lattice. The annihilation (creation) operator $C_r$ ($C^{\dagger}_r$) consists of a two-orbital and a spin-1/2 degrees of freedom.  We note that on-site disorder breaks the particle-hole symmetry.

Generally, $\alpha=\beta=1$ refers to the 2D isotropic (diagonally oriented) quasi-periodicity. 
We also consider purely longitudinal (transverse) quasi-periodicity only along $x$ ($y$)-direction choosing 
$\alpha=1,~\beta=0$ ($\alpha=0,~\beta=1$). The spatial configurations of these disorder potentials are demonstrated in Fig.~\ref{system} (b), (c), and (d) for $\phi=0$. Here, we consider a thickness of three quintuple layers such that the 
model becomes trivial in the clean and undoped limit with appropriate material parameters: $v_F=3.07$ eV\AA, $m_0=44$ meV, $B=37.3$ eV\AA$^2$ \cite{wang2015}, and $a=20$\AA\;\cite{Okugawa20}.
We compute the disorder averaged conductance $G$ (in units of $e^2/h$) for the central region and the corresponding standard deviation $\delta G$ (in units of $e^2/h$), following the Landauer-B\"uttiker formalism~\cite{Landauer1970, Buttiker1988} with recursive Green's function technique \cite{Rotter00,Rotter03,Libisch_2012,Okugawa20}, as a function of  both  disorder strength $W$ and exchange field $gM$ (see Figs.~\ref{xy_correlated}, \ref{x_correlated} and \ref{y_correlated}).  The  QSHI and  QSCI phases both are identified by quantized conductance $G=2$ (green) while the former [latter] appears in the absence [presence] of exchange field. The QAHI phases are characterized by quantized conductance $G=1$ (orange).

We also analyze the emergence of disorder mediated TPTs using the SCBA \cite{Groth2009}. 
Importantly, 
TAI phases appear when 
the renormalized topological mass $\overline {m}=m+\delta m$ becomes negative  $ \overline {m}<0$ as well as the renormalized chemical potential $\overline{E}_F=E_F +\delta \mu$ lies inside the band gap  $ |\overline {E}_F|<-\bar {m}$. 
Exploiting the block diagonal form of the Hamiltonian (Eq.~(\ref{hamiltonian})), we can decompose the self energy into upper and lower blocks  $\Sigma_{u,l}=\Sigma_{0}^{u,l}\sigma_{0}+\Sigma_{x}^{u,l}\sigma_{x}+\Sigma_{y}^{u,l}\sigma_{y}+\Sigma_{z}^{u,l}\sigma_{z}$. The  correction terms, caused by the disorder, are thus found to be 
$\delta m_{u,l}= {\rm Re}[\Sigma_{z}^{u,l}]$ and $\delta \mu_{u,l}= -{\rm Re}[\Sigma_{0}^{u,l}]$.
The self energy  can be expressed through self-consistent equations by incorporating $C(\bm k)$ i.e., Fourier transform of the
disorder correlation function $C_{m,n}$, as follows \cite{Zimmermann09,Girschik13,Fu21,SuppMater}
\begin{align}
\Sigma_{u,l}&= \int d{\bm k}~ C(\bm k)~(E_F + i \zeta - H_{u,l}(\bm k)-\Sigma_{u,l})^{-1}
\nonumber \\
&=W^2 (\Sigma^+_{u,l} + \Sigma^-_{u,l} )/16  
\label{SCBA_CPA} 
\end{align}
with $\Sigma^{\pm}_{u,l}=\left(E_F + i \zeta - H_{u,l}(\pm \alpha Q, \pm\beta Q)-\Sigma_{u,l}   \right)^{-1}$,
$Q=2 \pi \eta$ and $\zeta \to 0$.  
The phase boundaries can be determined by  $|\overline{E}_F^{u,l}|=-\overline{m}_{0}^{u,l}$ for $\overline{m}_{0}^{u,l}<0$ and  $|\overline{E}_F^{u,l}|=\overline{m}_{0}^{u,l}$ for $\overline{m}_{0}^{u,l}>0$  segregating
the TAI phases
with quantized $G\ne 0$ from the trivial phases with  non-quantized  $G$. Note that the SCBA 
fails to detect the TPTs to an AI phase for strong disorder $W>{\mathcal O}(B/a^2,v_F/a)$.

\textcolor{blue}{\textit{Results}---}
We start our discussion on TPTs induced by the correlated disorder with  Fig.~\ref{xy_correlated} where we consider the isotropic quasi-periodicity i.e., $\alpha=\beta=1$ (see Fig.~\ref{system} (a)) and compare to the random disorder case \cite{Okugawa20}.
The bulk gap $\Delta$ of the central system can be read off by the Fermi energy $E_F$ at which the quantized conductance $G=0,~1$, and $2$ (accompanied by $\delta G= 0$) can either change to a non-quantized value with  $\delta G\ne 0$ or a quantized value with $G>2$. The later phase with $G>2$ is exclusively observed for the case of correlated disorder which stems from extended bulk modes lying well above the trivial and topological gap.
The thin film is in the NI phase for  $\Delta=m_0= 44$ meV in the clean non-magnetic case (see Fig.~\ref{xy_correlated} (a)).  
With increasing disorder strength regardless of whether the disorder is correlated or uncorrelated, the trivial gap reduces and eventually the gap becomes topological at $(W,E_F)\approx(500~\rm{meV},0~\rm{meV})$, i.e., $\Delta<0$ due to band inversion. At this point the thin film enters into a QSHI phase with quantized  $G=2$. The QSHI phase has a maximal gap at $W\approx 900$ meV and for larger disorder $W$ rapidly turns into an AI. 
Upon including a magnetic field $gM=30 $ meV as shown in Fig.~\ref{xy_correlated} (e), the trivial gap $\Delta$ reduces to $14$ meV in the clean limit. The trivial system  now first enters into a QAHI    with quantized  $G=1$ upon inclusion of disorder, followed by a QSCI  phase with quantized $G=2$ and eventually an AI phase takes over for strong disorder. Upon increasing the magnetic field to $gM=52$ meV, as shown in Fig.~\ref{xy_correlated} (i), the system already resides in the QAHI phase, with topological gap $|\Delta|=8$ meV, even in the clean limit. For increasing disorder $W$, the system similarly traverses through a series of QAHI $\to$ QSCI $\to $ AI phases. However,  the size of the QSCI (QAHI) phase decreases (increases) significantly for $gM=52$ meV as compared to that of $gM=30 $meV.

The different TPTs  except  the transition to the AI phase at large disorder are well captured by the SCBA as indicated by the lines in Fig.~\ref{xy_correlated}.  The evolution of the bulk gap for the central system in various phases and their boundaries are schematically demonstrated in Figs. ~\ref{xy_correlated} (A)-(F).
All these above features, obtained for correlated disorder, are qualitatively similar to random disorder.
 However, the correlated disorder is found to stabilize the TAI phases more clearly than random disorder as evident from the standard deviation $\delta G$ profiles (see  Figs.~\ref{xy_correlated} (b), (d), (f), (h), (j), and (l))

For longitudinal quasi-periodicity $\alpha=1$ and $\beta=0$ (see Fig.~\ref{system} (b)) we find a qualitatively similar picture, as shown in  Fig.~\ref{x_correlated},  compared to the previous isotropic case. Interestingly, the marked differences are that the  quantized transport from the bulk mode as well as the  QSCI phase at larger magnetic fields $gM$ are absent in the present case. The latter is due to the reservoir effect: when the topological gap of the lead is less than or comparable to the gap of the central system, a hybridization of the edge modes in the central system with the bulk modes in the leads may occur. The reservoir effect is analyzed in more detail in the Sec. II of SM \cite{SuppMater}.

Next we analyze transverse quasi-periodicity along $y$-direction i.e., perpendicular to the transport direction, with $\alpha=0$ and $\beta=1$ (see Fig.~\ref{system} (c)).  Contrary to the above cases, the QSHI and QSCI phases continue to exist with a topological gap $\Delta<0$
even for strong disorder $W>1000$ meV irrespective of the values of exchange field  as shown in Figs.~
\ref{y_correlated} (a), (c) and (e). This effect unique to the case of transverse  quasi-periodicity can be captured within a simple SCBA approach. {It is noteworthy that the phase transition boundaries, evaluated by SCBA, are exactly the same in Figs.~\ref{x_correlated} and \ref{y_correlated}. This is due to the underlying $C_4$ symmetry of the clean central system. The above crystalline symmetry further ensures that the phase diagram for longitudinal (Figs.~\ref{x_correlated}) and transverse (Figs.~\ref{y_correlated}) quasi-periodicities would be interchanged once the leads are connected to the top and bottom instead of the left and right of the central system. 
The results only depend on the relative orientation of the quasi-periodicity with regard to the transport direction. 
}

\textcolor{blue}{\textit{Discussions}---}
It is evident from the above investigations that the upper and lower block of the Hamiltonian in Eq.~(\ref{hamiltonian}), 
can be selectively made topological under the appropriate  orientations of correlated disorder. 
With increasing disorder $W$, the lower block  becomes topological first followed by the upper block. 
As long as the direction of the magnetic field is not altered, the above  feature is observed in all the  cases (flipping the magnetic field flips upper and lower block). 
We note that the individual phases for $E_F>0$ would  also symmetrically appear for $E_F<0$ owing to the emergent spectral symmetry of the system under $W \to -W$ (see SM Sec. IV for more discussion) \cite{Okugawa20,SuppMater}.

The quasi-periodicity generically introduces mobility edges such  that there exist extended bulk states within an energy interval between $[E_-,E_+]$ in the middle of the single particle spectrum around zero energy (see SM Sec. V for more discussion) \cite{SuppMater}. One expects these states to participate in the electron transport through the central disordered  
system above its bulk gap $\Delta$ provided $|\Delta|<|E_+-E_-|$.
For the isotropic quasi-periodicity, this mechanism of bulk transport might lead to the quantized conductance with $G>2$ and $\delta G \to 0$ even though  $\overline{m}_{0}^{u,l}>0$ (see Fig.~\ref{xy_correlated}(a), (e) and (i)). These regions
appear just outside the NI and TAI phases when $E_-<E_F<E_+$ and $|E_F|>|\Delta|$, in marked contrast to the random disorder case.
The universality classes of the TPTs 
between the TAI and non-TAI phases,  \cite{refId0,Hashimoto08,Huckestein95,Girschik13,Fu21} reported here are left for future research.

On the other hand, for anisotropic quasi-periodicities, the mobility edge energy interval $|E_+-E_-|$ for the extended bulk states shrinks significantly resulting in the suppression of quantized transport from the bulk states lying above $|\Delta|$.  This could be the reason why non-quantized bulk conductance with $G>2$ and $\delta G> 0$ for longitudinal and  transverse  quasi-periodicities is observed. The edge transport along $x$-direction is severely (minimally) influenced for longitudinal (transverse) quasi-periodicity as the mid-gap conducting edge modes can (can not) become localized which are otherwise delocalized  along the transport direction at $y=1, L_y$ (see Figs.~\ref{system} (e)-(g)). 
Combining these insights one notes that the TAI phase diagram for isotropic quasi-periodicity is an admixture of the anisotropic quasi-periodicities along longitudinal and transverse directions.

Our findings suggest that in terms of the current transport, the AI phase does not emerge for the transverse quasi-periodic case unlike to the remaining  cases. In the strong disorder regime, localized states reside in the interior bulk of the central system. For the 1D quasi-periodicity, such localization is expected to occur only over the 1D line of lattice sites on which the quasi-periodicity is embedded. This further indicates that longitudinal quasi-periodicity leads to spatially separated localized states through which the transport current can not propagate. The same is also true for diagonal quasi-periodicity.  On the other hand, for transverse quasi-periodicity, such localization along the $y$-direction essentially allows the current flow. 


Our study is potentially relevant to model the experimental findings on QAHI phases in magnetic TIs \cite{chang2013experimental,Checkelsky2014Trajectory,Chang2016Obse,tokura2019magnetic,watanabe2019quantum,Satake20}. Besides solid state systems, TIs are also realized for ultracold atomic gases in optical lattices \cite{cooper11,Aidelsburger13,atala2014observation,aidelsburger2015measuring,jotzu2014experimental,flaschner2016experimental}. To be specific, The Su-Schrieffer-Heeger model \cite{SSH_79} has already been implemented in optical lattice \cite{atala2013direct}. Furthermore, the quasi-periodic potential has been proposed \cite{Guidoni97} and implemented \cite{roati2008anderson} in optical lattices. The optical laser speckle potential could be engineered to introduce the correlated disorder of the type discussed here \cite{billy2008direct,Lye05}. In the light of the above considerations, we believe that TPTs induced by the interplay between the correlated disorder and magnetism can be investigated with ultracold atoms.
\subsection*{Acknowledgments}
This work was supported by the Deutsche Forschungsgemeinschaft (DFG, German Research Foundation) via RTG 1995 and Germany’s Excellence Strategy - Cluster of Excellence Matter and Light for Quantum Computing (ML4Q) EXC 2004/1 - 390534769. Simulations were performed with computing resources granted by RWTH Aachen University under project rwth0857. We acknowledge support from the Max Planck-New York City Center for Non-Equilibrium Quantum Phenomena.

\subsection*{Author Contribution}
TO performed the calculations. TN conceived the project. TN and DMK supervised the project. All the authors contributed to the manuscript.




\appendix

\newcounter{defcounter}
\setcounter{defcounter}{0}




\vspace{20pt}

\bibliography{Manuscript}{}

\providecommand{\noopsort}[1]{}\providecommand{\singleletter}[1]{#1}%
\begin{thebibliography}{98}%
\makeatletter
\providecommand \@ifxundefined [1]{%
 \@ifx{#1\undefined}
}%
\providecommand \@ifnum [1]{%
 \ifnum #1\expandafter \@firstoftwo
 \else \expandafter \@secondoftwo
 \fi
}%
\providecommand \@ifx [1]{%
 \ifx #1\expandafter \@firstoftwo
 \else \expandafter \@secondoftwo
 \fi
}%
\providecommand \natexlab [1]{#1}%
\providecommand \enquote  [1]{``#1''}%
\providecommand \bibnamefont  [1]{#1}%
\providecommand \bibfnamefont [1]{#1}%
\providecommand \citenamefont [1]{#1}%
\providecommand \href@noop [0]{\@secondoftwo}%
\providecommand \href [0]{\begingroup \@sanitize@url \@href}%
\providecommand \@href[1]{\@@startlink{#1}\@@href}%
\providecommand \@@href[1]{\endgroup#1\@@endlink}%
\providecommand \@sanitize@url [0]{\catcode `\\12\catcode `\$12\catcode
  `\&12\catcode `\#12\catcode `\^12\catcode `\_12\catcode `\%12\relax}%
\providecommand \@@startlink[1]{}%
\providecommand \@@endlink[0]{}%
\providecommand \url  [0]{\begingroup\@sanitize@url \@url }%
\providecommand \@url [1]{\endgroup\@href {#1}{\urlprefix }}%
\providecommand \urlprefix  [0]{URL }%
\providecommand \Eprint [0]{\href }%
\providecommand \doibase [0]{http://dx.doi.org/}%
\providecommand \selectlanguage [0]{\@gobble}%
\providecommand \bibinfo  [0]{\@secondoftwo}%
\providecommand \bibfield  [0]{\@secondoftwo}%
\providecommand \translation [1]{[#1]}%
\providecommand \BibitemOpen [0]{}%
\providecommand \bibitemStop [0]{}%
\providecommand \bibitemNoStop [0]{.\EOS\space}%
\providecommand \EOS [0]{\spacefactor3000\relax}%
\providecommand \BibitemShut  [1]{\csname bibitem#1\endcsname}%
\let\auto@bib@innerbib\@empty
\bibitem [{\citenamefont {Haldane}(1988)}]{Haldane88}%
  \BibitemOpen
  \bibfield  {author} {\bibinfo {author} {\bibfnamefont {F.~D.~M.}\
  \bibnamefont {Haldane}},\ }\href {\doibase 10.1103/PhysRevLett.61.2015}
  {\bibfield  {journal} {\bibinfo  {journal} {Phys. Rev. Lett.}\ }\textbf
  {\bibinfo {volume} {61}},\ \bibinfo {pages} {2015} (\bibinfo {year}
  {1988})}\BibitemShut {NoStop}%
\bibitem [{\citenamefont {Onoda}\ and\ \citenamefont
  {Nagaosa}(2003)}]{Onoda03}%
  \BibitemOpen
  \bibfield  {author} {\bibinfo {author} {\bibfnamefont {M.}~\bibnamefont
  {Onoda}}\ and\ \bibinfo {author} {\bibfnamefont {N.}~\bibnamefont
  {Nagaosa}},\ }\href {\doibase 10.1103/PhysRevLett.90.206601} {\bibfield
  {journal} {\bibinfo  {journal} {Phys. Rev. Lett.}\ }\textbf {\bibinfo
  {volume} {90}},\ \bibinfo {pages} {206601} (\bibinfo {year}
  {2003})}\BibitemShut {NoStop}%
\bibitem [{\citenamefont {Kane}\ and\ \citenamefont {Mele}(2005)}]{Kane05a}%
  \BibitemOpen
  \bibfield  {author} {\bibinfo {author} {\bibfnamefont {C.~L.}\ \bibnamefont
  {Kane}}\ and\ \bibinfo {author} {\bibfnamefont {E.~J.}\ \bibnamefont
  {Mele}},\ }\href {\doibase 10.1103/PhysRevLett.95.226801} {\bibfield
  {journal} {\bibinfo  {journal} {Phys. Rev. Lett.}\ }\textbf {\bibinfo
  {volume} {95}},\ \bibinfo {pages} {226801} (\bibinfo {year}
  {2005})}\BibitemShut {NoStop}%
\bibitem [{\citenamefont {Sheng}\ \emph {et~al.}(2006)\citenamefont {Sheng},
  \citenamefont {Weng}, \citenamefont {Sheng},\ and\ \citenamefont
  {Haldane}}]{ShengDN2006}%
  \BibitemOpen
  \bibfield  {author} {\bibinfo {author} {\bibfnamefont {D.~N.}\ \bibnamefont
  {Sheng}}, \bibinfo {author} {\bibfnamefont {Z.~Y.}\ \bibnamefont {Weng}},
  \bibinfo {author} {\bibfnamefont {L.}~\bibnamefont {Sheng}}, \ and\ \bibinfo
  {author} {\bibfnamefont {F.~D.~M.}\ \bibnamefont {Haldane}},\ }\href
  {\doibase 10.1103/PhysRevLett.97.036808} {\bibfield  {journal} {\bibinfo
  {journal} {Phys. Rev. Lett.}\ }\textbf {\bibinfo {volume} {97}},\ \bibinfo
  {pages} {036808} (\bibinfo {year} {2006})}\BibitemShut {NoStop}%
\bibitem [{\citenamefont {Bernevig}\ and\ \citenamefont
  {Zhang}(2006)}]{Bernevig06}%
  \BibitemOpen
  \bibfield  {author} {\bibinfo {author} {\bibfnamefont {B.~A.}\ \bibnamefont
  {Bernevig}}\ and\ \bibinfo {author} {\bibfnamefont {S.-C.}\ \bibnamefont
  {Zhang}},\ }\href {\doibase 10.1103/PhysRevLett.96.106802} {\bibfield
  {journal} {\bibinfo  {journal} {Phys. Rev. Lett.}\ }\textbf {\bibinfo
  {volume} {96}},\ \bibinfo {pages} {106802} (\bibinfo {year}
  {2006})}\BibitemShut {NoStop}%
\bibitem [{\citenamefont {Hasan}\ and\ \citenamefont {Kane}(2010)}]{Hasan10}%
  \BibitemOpen
  \bibfield  {author} {\bibinfo {author} {\bibfnamefont {M.~Z.}\ \bibnamefont
  {Hasan}}\ and\ \bibinfo {author} {\bibfnamefont {C.~L.}\ \bibnamefont
  {Kane}},\ }\href {\doibase 10.1103/RevModPhys.82.3045} {\bibfield  {journal}
  {\bibinfo  {journal} {Rev. Mod. Phys.}\ }\textbf {\bibinfo {volume} {82}},\
  \bibinfo {pages} {3045} (\bibinfo {year} {2010})}\BibitemShut {NoStop}%
\bibitem [{\citenamefont {Liu}\ \emph {et~al.}(2016{\natexlab{a}})\citenamefont
  {Liu}, \citenamefont {Zhang},\ and\ \citenamefont {Qi}}]{Liu16}%
  \BibitemOpen
  \bibfield  {author} {\bibinfo {author} {\bibfnamefont {C.-X.}\ \bibnamefont
  {Liu}}, \bibinfo {author} {\bibfnamefont {S.-C.}\ \bibnamefont {Zhang}}, \
  and\ \bibinfo {author} {\bibfnamefont {X.-L.}\ \bibnamefont {Qi}},\ }\href
  {\doibase 10.1146/annurev-conmatphys-031115-011417} {\bibfield  {journal}
  {\bibinfo  {journal} {Annual Review of Condensed Matter Physics}\ }\textbf
  {\bibinfo {volume} {7}},\ \bibinfo {pages} {301} (\bibinfo {year}
  {2016}{\natexlab{a}})},\ \Eprint
  {http://arxiv.org/abs/https://doi.org/10.1146/annurev-conmatphys-031115-011417}
  {https://doi.org/10.1146/annurev-conmatphys-031115-011417} \BibitemShut
  {NoStop}%
\bibitem [{\citenamefont {Armitage}\ \emph {et~al.}(2018)\citenamefont
  {Armitage}, \citenamefont {Mele},\ and\ \citenamefont
  {Vishwanath}}]{Armitage2018}%
  \BibitemOpen
  \bibfield  {author} {\bibinfo {author} {\bibfnamefont {N.~P.}\ \bibnamefont
  {Armitage}}, \bibinfo {author} {\bibfnamefont {E.~J.}\ \bibnamefont {Mele}},
  \ and\ \bibinfo {author} {\bibfnamefont {A.}~\bibnamefont {Vishwanath}},\
  }\href {\doibase 10.1103/RevModPhys.90.015001} {\bibfield  {journal}
  {\bibinfo  {journal} {Rev. Mod. Phys.}\ }\textbf {\bibinfo {volume} {90}},\
  \bibinfo {pages} {015001} (\bibinfo {year} {2018})}\BibitemShut {NoStop}%
\bibitem [{\citenamefont {Bernevig}\ \emph {et~al.}(2006)\citenamefont
  {Bernevig}, \citenamefont {Hughes},\ and\ \citenamefont
  {Zhang}}]{Bernevig1757}%
  \BibitemOpen
  \bibfield  {author} {\bibinfo {author} {\bibfnamefont {B.~A.}\ \bibnamefont
  {Bernevig}}, \bibinfo {author} {\bibfnamefont {T.~L.}\ \bibnamefont
  {Hughes}}, \ and\ \bibinfo {author} {\bibfnamefont {S.-C.}\ \bibnamefont
  {Zhang}},\ }\href {\doibase 10.1126/science.1133734} {\bibfield  {journal}
  {\bibinfo  {journal} {Science}\ }\textbf {\bibinfo {volume} {314}},\ \bibinfo
  {pages} {1757} (\bibinfo {year} {2006})}\BibitemShut {NoStop}%
\bibitem [{\citenamefont {Zhang}\ \emph {et~al.}(2009)\citenamefont {Zhang},
  \citenamefont {Liu}, \citenamefont {Qi}, \citenamefont {Dai}, \citenamefont
  {Fang},\ and\ \citenamefont {Zhang}}]{zhang2009topological}%
  \BibitemOpen
  \bibfield  {author} {\bibinfo {author} {\bibfnamefont {H.}~\bibnamefont
  {Zhang}}, \bibinfo {author} {\bibfnamefont {C.-X.}\ \bibnamefont {Liu}},
  \bibinfo {author} {\bibfnamefont {X.-L.}\ \bibnamefont {Qi}}, \bibinfo
  {author} {\bibfnamefont {X.}~\bibnamefont {Dai}}, \bibinfo {author}
  {\bibfnamefont {Z.}~\bibnamefont {Fang}}, \ and\ \bibinfo {author}
  {\bibfnamefont {S.-C.}\ \bibnamefont {Zhang}},\ }\href@noop {} {\bibfield
  {journal} {\bibinfo  {journal} {Nature physics}\ }\textbf {\bibinfo {volume}
  {5}},\ \bibinfo {pages} {438} (\bibinfo {year} {2009})}\BibitemShut {NoStop}%
\bibitem [{\citenamefont {K{\"o}nig}\ \emph {et~al.}(2007)\citenamefont
  {K{\"o}nig}, \citenamefont {Wiedmann}, \citenamefont {Br{\"u}ne},
  \citenamefont {Roth}, \citenamefont {Buhmann}, \citenamefont {Molenkamp},
  \citenamefont {Qi},\ and\ \citenamefont {Zhang}}]{konig2007quantum}%
  \BibitemOpen
  \bibfield  {author} {\bibinfo {author} {\bibfnamefont {M.}~\bibnamefont
  {K{\"o}nig}}, \bibinfo {author} {\bibfnamefont {S.}~\bibnamefont {Wiedmann}},
  \bibinfo {author} {\bibfnamefont {C.}~\bibnamefont {Br{\"u}ne}}, \bibinfo
  {author} {\bibfnamefont {A.}~\bibnamefont {Roth}}, \bibinfo {author}
  {\bibfnamefont {H.}~\bibnamefont {Buhmann}}, \bibinfo {author} {\bibfnamefont
  {L.~W.}\ \bibnamefont {Molenkamp}}, \bibinfo {author} {\bibfnamefont {X.-L.}\
  \bibnamefont {Qi}}, \ and\ \bibinfo {author} {\bibfnamefont {S.-C.}\
  \bibnamefont {Zhang}},\ }\href@noop {} {\bibfield  {journal} {\bibinfo
  {journal} {Science}\ }\textbf {\bibinfo {volume} {318}},\ \bibinfo {pages}
  {766} (\bibinfo {year} {2007})}\BibitemShut {NoStop}%
\bibitem [{\citenamefont {Roth}\ \emph {et~al.}(2009)\citenamefont {Roth},
  \citenamefont {Br{\"u}ne}, \citenamefont {Buhmann}, \citenamefont
  {Molenkamp}, \citenamefont {Maciejko}, \citenamefont {Qi},\ and\
  \citenamefont {Zhang}}]{roth2009nonlocal}%
  \BibitemOpen
  \bibfield  {author} {\bibinfo {author} {\bibfnamefont {A.}~\bibnamefont
  {Roth}}, \bibinfo {author} {\bibfnamefont {C.}~\bibnamefont {Br{\"u}ne}},
  \bibinfo {author} {\bibfnamefont {H.}~\bibnamefont {Buhmann}}, \bibinfo
  {author} {\bibfnamefont {L.~W.}\ \bibnamefont {Molenkamp}}, \bibinfo {author}
  {\bibfnamefont {J.}~\bibnamefont {Maciejko}}, \bibinfo {author}
  {\bibfnamefont {X.-L.}\ \bibnamefont {Qi}}, \ and\ \bibinfo {author}
  {\bibfnamefont {S.-C.}\ \bibnamefont {Zhang}},\ }\href@noop {} {\bibfield
  {journal} {\bibinfo  {journal} {Science}\ }\textbf {\bibinfo {volume}
  {325}},\ \bibinfo {pages} {294} (\bibinfo {year} {2009})}\BibitemShut
  {NoStop}%
\bibitem [{\citenamefont {Mong}\ \emph {et~al.}(2010)\citenamefont {Mong},
  \citenamefont {Essin},\ and\ \citenamefont {Moore}}]{Mong2010}%
  \BibitemOpen
  \bibfield  {author} {\bibinfo {author} {\bibfnamefont {R.~S.~K.}\
  \bibnamefont {Mong}}, \bibinfo {author} {\bibfnamefont {A.~M.}\ \bibnamefont
  {Essin}}, \ and\ \bibinfo {author} {\bibfnamefont {J.~E.}\ \bibnamefont
  {Moore}},\ }\href {\doibase 10.1103/PhysRevB.81.245209} {\bibfield  {journal}
  {\bibinfo  {journal} {Phys. Rev. B}\ }\textbf {\bibinfo {volume} {81}},\
  \bibinfo {pages} {245209} (\bibinfo {year} {2010})}\BibitemShut {NoStop}%
\bibitem [{\citenamefont {Otrokov}\ \emph {et~al.}(2019)\citenamefont
  {Otrokov}, \citenamefont {Klimovskikh}, \citenamefont {Bentmann},
  \citenamefont {Estyunin}, \citenamefont {Zeugner}, \citenamefont {Aliev},
  \citenamefont {Ga{\ss}}, \citenamefont {Wolter}, \citenamefont {Koroleva},
  \citenamefont {Shikin} \emph {et~al.}}]{otrokov2019prediction}%
  \BibitemOpen
  \bibfield  {author} {\bibinfo {author} {\bibfnamefont {M.~M.}\ \bibnamefont
  {Otrokov}}, \bibinfo {author} {\bibfnamefont {I.~I.}\ \bibnamefont
  {Klimovskikh}}, \bibinfo {author} {\bibfnamefont {H.}~\bibnamefont
  {Bentmann}}, \bibinfo {author} {\bibfnamefont {D.}~\bibnamefont {Estyunin}},
  \bibinfo {author} {\bibfnamefont {A.}~\bibnamefont {Zeugner}}, \bibinfo
  {author} {\bibfnamefont {Z.~S.}\ \bibnamefont {Aliev}}, \bibinfo {author}
  {\bibfnamefont {S.}~\bibnamefont {Ga{\ss}}}, \bibinfo {author} {\bibfnamefont
  {A.}~\bibnamefont {Wolter}}, \bibinfo {author} {\bibfnamefont
  {A.}~\bibnamefont {Koroleva}}, \bibinfo {author} {\bibfnamefont {A.~M.}\
  \bibnamefont {Shikin}},  \emph {et~al.},\ }\href {\doibase
  10.1038/s41586-019-1840-9} {\bibfield  {journal} {\bibinfo  {journal}
  {Nature}\ }\textbf {\bibinfo {volume} {576}},\ \bibinfo {pages} {416}
  (\bibinfo {year} {2019})}\BibitemShut {NoStop}%
\bibitem [{\citenamefont {Li}\ \emph {et~al.}(2019)\citenamefont {Li},
  \citenamefont {Li}, \citenamefont {Du}, \citenamefont {Wang}, \citenamefont
  {Gu}, \citenamefont {Zhang}, \citenamefont {He}, \citenamefont {Duan},\ and\
  \citenamefont {Xu}}]{liJH2018}%
  \BibitemOpen
  \bibfield  {author} {\bibinfo {author} {\bibfnamefont {J.}~\bibnamefont
  {Li}}, \bibinfo {author} {\bibfnamefont {Y.}~\bibnamefont {Li}}, \bibinfo
  {author} {\bibfnamefont {S.}~\bibnamefont {Du}}, \bibinfo {author}
  {\bibfnamefont {Z.}~\bibnamefont {Wang}}, \bibinfo {author} {\bibfnamefont
  {B.-L.}\ \bibnamefont {Gu}}, \bibinfo {author} {\bibfnamefont {S.-C.}\
  \bibnamefont {Zhang}}, \bibinfo {author} {\bibfnamefont {K.}~\bibnamefont
  {He}}, \bibinfo {author} {\bibfnamefont {W.}~\bibnamefont {Duan}}, \ and\
  \bibinfo {author} {\bibfnamefont {Y.}~\bibnamefont {Xu}},\ }\href@noop {}
  {\bibfield  {journal} {\bibinfo  {journal} {Sci. Adv.}\ }\textbf {\bibinfo
  {volume} {\textbf{5}}},\ \bibinfo {pages} {eaaw5685} (\bibinfo {year}
  {2019})}\BibitemShut {NoStop}%
\bibitem [{\citenamefont {Tang}\ \emph {et~al.}(2016)\citenamefont {Tang},
  \citenamefont {Zhou}, \citenamefont {Xu},\ and\ \citenamefont
  {Zhang}}]{tang2016dirac}%
  \BibitemOpen
  \bibfield  {author} {\bibinfo {author} {\bibfnamefont {P.}~\bibnamefont
  {Tang}}, \bibinfo {author} {\bibfnamefont {Q.}~\bibnamefont {Zhou}}, \bibinfo
  {author} {\bibfnamefont {G.}~\bibnamefont {Xu}}, \ and\ \bibinfo {author}
  {\bibfnamefont {S.-C.}\ \bibnamefont {Zhang}},\ }\href@noop {} {\bibfield
  {journal} {\bibinfo  {journal} {Nat. Phys.}\ }\textbf {\bibinfo {volume}
  {\textbf{12}}},\ \bibinfo {pages} {1100} (\bibinfo {year}
  {2016})}\BibitemShut {NoStop}%
\bibitem [{\citenamefont {Liu}\ \emph {et~al.}(2016{\natexlab{b}})\citenamefont
  {Liu}, \citenamefont {Zhang},\ and\ \citenamefont {Qi}}]{liu2016quantum}%
  \BibitemOpen
  \bibfield  {author} {\bibinfo {author} {\bibfnamefont {C.-X.}\ \bibnamefont
  {Liu}}, \bibinfo {author} {\bibfnamefont {S.-C.}\ \bibnamefont {Zhang}}, \
  and\ \bibinfo {author} {\bibfnamefont {X.-L.}\ \bibnamefont {Qi}},\ }\href
  {\doibase 10.1146/annurev-conmatphys-031115-011417} {\bibfield  {journal}
  {\bibinfo  {journal} {Annu. Rev. Condens. Matter Phys.}\ }\textbf {\bibinfo
  {volume} {7}},\ \bibinfo {pages} {301} (\bibinfo {year}
  {2016}{\natexlab{b}})}\BibitemShut {NoStop}%
\bibitem [{\citenamefont {He}\ \emph {et~al.}(2018)\citenamefont {He},
  \citenamefont {Wang},\ and\ \citenamefont {Xue}}]{he2018topological}%
  \BibitemOpen
  \bibfield  {author} {\bibinfo {author} {\bibfnamefont {K.}~\bibnamefont
  {He}}, \bibinfo {author} {\bibfnamefont {Y.}~\bibnamefont {Wang}}, \ and\
  \bibinfo {author} {\bibfnamefont {Q.-K.}\ \bibnamefont {Xue}},\ }\href
  {\doibase 10.1146/annurev-conmatphys-033117-054144} {\bibfield  {journal}
  {\bibinfo  {journal} {Annu. Rev. Condens. Matter Phys.}\ }\textbf {\bibinfo
  {volume} {9}},\ \bibinfo {pages} {329} (\bibinfo {year} {2018})}\BibitemShut
  {NoStop}%
\bibitem [{\citenamefont {Saha}\ \emph {et~al.}(2021)\citenamefont {Saha},
  \citenamefont {Nag},\ and\ \citenamefont {Mandal}}]{Saha21}%
  \BibitemOpen
  \bibfield  {author} {\bibinfo {author} {\bibfnamefont {S.}~\bibnamefont
  {Saha}}, \bibinfo {author} {\bibfnamefont {T.}~\bibnamefont {Nag}}, \ and\
  \bibinfo {author} {\bibfnamefont {S.}~\bibnamefont {Mandal}},\ }\href
  {\doibase 10.1103/PhysRevB.103.235154} {\bibfield  {journal} {\bibinfo
  {journal} {Phys. Rev. B}\ }\textbf {\bibinfo {volume} {103}},\ \bibinfo
  {pages} {235154} (\bibinfo {year} {2021})}\BibitemShut {NoStop}%
\bibitem [{\citenamefont {Liu}\ \emph {et~al.}(2008)\citenamefont {Liu},
  \citenamefont {Qi}, \citenamefont {Dai}, \citenamefont {Fang},\ and\
  \citenamefont {Zhang}}]{Liu08}%
  \BibitemOpen
  \bibfield  {author} {\bibinfo {author} {\bibfnamefont {C.-X.}\ \bibnamefont
  {Liu}}, \bibinfo {author} {\bibfnamefont {X.-L.}\ \bibnamefont {Qi}},
  \bibinfo {author} {\bibfnamefont {X.}~\bibnamefont {Dai}}, \bibinfo {author}
  {\bibfnamefont {Z.}~\bibnamefont {Fang}}, \ and\ \bibinfo {author}
  {\bibfnamefont {S.-C.}\ \bibnamefont {Zhang}},\ }\href {\doibase
  10.1103/PhysRevLett.101.146802} {\bibfield  {journal} {\bibinfo  {journal}
  {Phys. Rev. Lett.}\ }\textbf {\bibinfo {volume} {101}},\ \bibinfo {pages}
  {146802} (\bibinfo {year} {2008})}\BibitemShut {NoStop}%
\bibitem [{\citenamefont {Li}\ \emph {et~al.}(2013)\citenamefont {Li},
  \citenamefont {Sheng}, \citenamefont {Shen}, \citenamefont {Shao},
  \citenamefont {Wang}, \citenamefont {Sheng},\ and\ \citenamefont
  {Xing}}]{Li13}%
  \BibitemOpen
  \bibfield  {author} {\bibinfo {author} {\bibfnamefont {H.}~\bibnamefont
  {Li}}, \bibinfo {author} {\bibfnamefont {L.}~\bibnamefont {Sheng}}, \bibinfo
  {author} {\bibfnamefont {R.}~\bibnamefont {Shen}}, \bibinfo {author}
  {\bibfnamefont {L.~B.}\ \bibnamefont {Shao}}, \bibinfo {author}
  {\bibfnamefont {B.}~\bibnamefont {Wang}}, \bibinfo {author} {\bibfnamefont
  {D.~N.}\ \bibnamefont {Sheng}}, \ and\ \bibinfo {author} {\bibfnamefont
  {D.~Y.}\ \bibnamefont {Xing}},\ }\href {\doibase
  10.1103/PhysRevLett.110.266802} {\bibfield  {journal} {\bibinfo  {journal}
  {Phys. Rev. Lett.}\ }\textbf {\bibinfo {volume} {110}},\ \bibinfo {pages}
  {266802} (\bibinfo {year} {2013})}\BibitemShut {NoStop}%
\bibitem [{\citenamefont {Qiao}\ \emph {et~al.}(2010)\citenamefont {Qiao},
  \citenamefont {Yang}, \citenamefont {Feng}, \citenamefont {Tse},
  \citenamefont {Ding}, \citenamefont {Yao}, \citenamefont {Wang},\ and\
  \citenamefont {Niu}}]{Qiao10}%
  \BibitemOpen
  \bibfield  {author} {\bibinfo {author} {\bibfnamefont {Z.}~\bibnamefont
  {Qiao}}, \bibinfo {author} {\bibfnamefont {S.~A.}\ \bibnamefont {Yang}},
  \bibinfo {author} {\bibfnamefont {W.}~\bibnamefont {Feng}}, \bibinfo {author}
  {\bibfnamefont {W.-K.}\ \bibnamefont {Tse}}, \bibinfo {author} {\bibfnamefont
  {J.}~\bibnamefont {Ding}}, \bibinfo {author} {\bibfnamefont {Y.}~\bibnamefont
  {Yao}}, \bibinfo {author} {\bibfnamefont {J.}~\bibnamefont {Wang}}, \ and\
  \bibinfo {author} {\bibfnamefont {Q.}~\bibnamefont {Niu}},\ }\href {\doibase
  10.1103/PhysRevB.82.161414} {\bibfield  {journal} {\bibinfo  {journal} {Phys.
  Rev. B}\ }\textbf {\bibinfo {volume} {82}},\ \bibinfo {pages} {161414}
  (\bibinfo {year} {2010})}\BibitemShut {NoStop}%
\bibitem [{\citenamefont {Yang}\ \emph {et~al.}(2011)\citenamefont {Yang},
  \citenamefont {Xu}, \citenamefont {Sheng}, \citenamefont {Wang},
  \citenamefont {Xing},\ and\ \citenamefont {Sheng}}]{Yang11}%
  \BibitemOpen
  \bibfield  {author} {\bibinfo {author} {\bibfnamefont {Y.}~\bibnamefont
  {Yang}}, \bibinfo {author} {\bibfnamefont {Z.}~\bibnamefont {Xu}}, \bibinfo
  {author} {\bibfnamefont {L.}~\bibnamefont {Sheng}}, \bibinfo {author}
  {\bibfnamefont {B.}~\bibnamefont {Wang}}, \bibinfo {author} {\bibfnamefont
  {D.~Y.}\ \bibnamefont {Xing}}, \ and\ \bibinfo {author} {\bibfnamefont
  {D.~N.}\ \bibnamefont {Sheng}},\ }\href {\doibase
  10.1103/PhysRevLett.107.066602} {\bibfield  {journal} {\bibinfo  {journal}
  {Phys. Rev. Lett.}\ }\textbf {\bibinfo {volume} {107}},\ \bibinfo {pages}
  {066602} (\bibinfo {year} {2011})}\BibitemShut {NoStop}%
\bibitem [{\citenamefont {Luo}\ \emph {et~al.}(2017)\citenamefont {Luo},
  \citenamefont {Shao}, \citenamefont {Deng}, \citenamefont {Deng},\ and\
  \citenamefont {Sheng}}]{luo2017time}%
  \BibitemOpen
  \bibfield  {author} {\bibinfo {author} {\bibfnamefont {W.}~\bibnamefont
  {Luo}}, \bibinfo {author} {\bibfnamefont {D.}~\bibnamefont {Shao}}, \bibinfo
  {author} {\bibfnamefont {M.-X.}\ \bibnamefont {Deng}}, \bibinfo {author}
  {\bibfnamefont {W.}~\bibnamefont {Deng}}, \ and\ \bibinfo {author}
  {\bibfnamefont {L.}~\bibnamefont {Sheng}},\ }\href@noop {} {\bibfield
  {journal} {\bibinfo  {journal} {Scientific reports}\ }\textbf {\bibinfo
  {volume} {7}},\ \bibinfo {pages} {43049} (\bibinfo {year}
  {2017})}\BibitemShut {NoStop}%
\bibitem [{\citenamefont {Yu}\ \emph {et~al.}(2010)\citenamefont {Yu},
  \citenamefont {Zhang}, \citenamefont {Zhang}, \citenamefont {Zhang},
  \citenamefont {Dai},\ and\ \citenamefont {Fang}}]{yu2010quantized}%
  \BibitemOpen
  \bibfield  {author} {\bibinfo {author} {\bibfnamefont {R.}~\bibnamefont
  {Yu}}, \bibinfo {author} {\bibfnamefont {W.}~\bibnamefont {Zhang}}, \bibinfo
  {author} {\bibfnamefont {H.-J.}\ \bibnamefont {Zhang}}, \bibinfo {author}
  {\bibfnamefont {S.-C.}\ \bibnamefont {Zhang}}, \bibinfo {author}
  {\bibfnamefont {X.}~\bibnamefont {Dai}}, \ and\ \bibinfo {author}
  {\bibfnamefont {Z.}~\bibnamefont {Fang}},\ }\href {\doibase
  10.1126/science.1187485} {\bibfield  {journal} {\bibinfo  {journal}
  {Science}\ }\textbf {\bibinfo {volume} {329}},\ \bibinfo {pages} {61}
  (\bibinfo {year} {2010})}\BibitemShut {NoStop}%
\bibitem [{\citenamefont {Chang}\ \emph {et~al.}(2013)\citenamefont {Chang},
  \citenamefont {Zhang}, \citenamefont {Feng}, \citenamefont {Shen},
  \citenamefont {Zhang}, \citenamefont {Guo}, \citenamefont {Li}, \citenamefont
  {Ou}, \citenamefont {Wei}, \citenamefont {Wang} \emph
  {et~al.}}]{chang2013experimental}%
  \BibitemOpen
  \bibfield  {author} {\bibinfo {author} {\bibfnamefont {C.-Z.}\ \bibnamefont
  {Chang}}, \bibinfo {author} {\bibfnamefont {J.}~\bibnamefont {Zhang}},
  \bibinfo {author} {\bibfnamefont {X.}~\bibnamefont {Feng}}, \bibinfo {author}
  {\bibfnamefont {J.}~\bibnamefont {Shen}}, \bibinfo {author} {\bibfnamefont
  {Z.}~\bibnamefont {Zhang}}, \bibinfo {author} {\bibfnamefont
  {M.}~\bibnamefont {Guo}}, \bibinfo {author} {\bibfnamefont {K.}~\bibnamefont
  {Li}}, \bibinfo {author} {\bibfnamefont {Y.}~\bibnamefont {Ou}}, \bibinfo
  {author} {\bibfnamefont {P.}~\bibnamefont {Wei}}, \bibinfo {author}
  {\bibfnamefont {L.-L.}\ \bibnamefont {Wang}},  \emph {et~al.},\ }\href
  {\doibase 10.1126/science.1234414} {\bibfield  {journal} {\bibinfo  {journal}
  {Science}\ }\textbf {\bibinfo {volume} {340}},\ \bibinfo {pages} {167}
  (\bibinfo {year} {2013})}\BibitemShut {NoStop}%
\bibitem [{\citenamefont {Checkelsky}\ \emph {et~al.}(2014)\citenamefont
  {Checkelsky}, \citenamefont {Yoshimi}, \citenamefont {Tsukazaki},
  \citenamefont {Takahashi}, \citenamefont {Kozuka}, \citenamefont {Falson},
  \citenamefont {Kawasaki},\ and\ \citenamefont
  {Tokura}}]{Checkelsky2014Trajectory}%
  \BibitemOpen
  \bibfield  {author} {\bibinfo {author} {\bibfnamefont {J.~G.}\ \bibnamefont
  {Checkelsky}}, \bibinfo {author} {\bibfnamefont {R.}~\bibnamefont {Yoshimi}},
  \bibinfo {author} {\bibfnamefont {A.}~\bibnamefont {Tsukazaki}}, \bibinfo
  {author} {\bibfnamefont {K.~S.}\ \bibnamefont {Takahashi}}, \bibinfo {author}
  {\bibfnamefont {Y.}~\bibnamefont {Kozuka}}, \bibinfo {author} {\bibfnamefont
  {J.}~\bibnamefont {Falson}}, \bibinfo {author} {\bibfnamefont
  {M.}~\bibnamefont {Kawasaki}}, \ and\ \bibinfo {author} {\bibfnamefont
  {Y.}~\bibnamefont {Tokura}},\ }\href {\doibase 10.1038/nphys3053} {\bibfield
  {journal} {\bibinfo  {journal} {Nat. Phys.}\ }\textbf {\bibinfo {volume}
  {10}},\ \bibinfo {pages} {731} (\bibinfo {year} {2014})}\BibitemShut
  {NoStop}%
\bibitem [{\citenamefont {Chang}\ \emph {et~al.}(2015)\citenamefont {Chang},
  \citenamefont {Zhao}, \citenamefont {Kim}, \citenamefont {Zhang},
  \citenamefont {Assaf}, \citenamefont {Heiman}, \citenamefont {Zhang},
  \citenamefont {Liu}, \citenamefont {Chan},\ and\ \citenamefont
  {Moodera}}]{chang2015high}%
  \BibitemOpen
  \bibfield  {author} {\bibinfo {author} {\bibfnamefont {C.-Z.}\ \bibnamefont
  {Chang}}, \bibinfo {author} {\bibfnamefont {W.}~\bibnamefont {Zhao}},
  \bibinfo {author} {\bibfnamefont {D.~Y.}\ \bibnamefont {Kim}}, \bibinfo
  {author} {\bibfnamefont {H.}~\bibnamefont {Zhang}}, \bibinfo {author}
  {\bibfnamefont {B.~A.}\ \bibnamefont {Assaf}}, \bibinfo {author}
  {\bibfnamefont {D.}~\bibnamefont {Heiman}}, \bibinfo {author} {\bibfnamefont
  {S.-C.}\ \bibnamefont {Zhang}}, \bibinfo {author} {\bibfnamefont
  {C.}~\bibnamefont {Liu}}, \bibinfo {author} {\bibfnamefont {M.~H.}\
  \bibnamefont {Chan}}, \ and\ \bibinfo {author} {\bibfnamefont {J.~S.}\
  \bibnamefont {Moodera}},\ }\href {\doibase 10.1038/nmat4204} {\bibfield
  {journal} {\bibinfo  {journal} {Nat. Mater.}\ }\textbf {\bibinfo {volume}
  {14}},\ \bibinfo {pages} {473} (\bibinfo {year} {2015})}\BibitemShut
  {NoStop}%
\bibitem [{\citenamefont {Feng}\ \emph {et~al.}(2015)\citenamefont {Feng},
  \citenamefont {Feng}, \citenamefont {Ou}, \citenamefont {Wang}, \citenamefont
  {Liu}, \citenamefont {Zhang}, \citenamefont {Zhao}, \citenamefont {Jiang},
  \citenamefont {Zhang}, \citenamefont {He}, \citenamefont {Ma}, \citenamefont
  {Xue},\ and\ \citenamefont {Wang}}]{FengYang2015}%
  \BibitemOpen
  \bibfield  {author} {\bibinfo {author} {\bibfnamefont {Y.}~\bibnamefont
  {Feng}}, \bibinfo {author} {\bibfnamefont {X.}~\bibnamefont {Feng}}, \bibinfo
  {author} {\bibfnamefont {Y.}~\bibnamefont {Ou}}, \bibinfo {author}
  {\bibfnamefont {J.}~\bibnamefont {Wang}}, \bibinfo {author} {\bibfnamefont
  {C.}~\bibnamefont {Liu}}, \bibinfo {author} {\bibfnamefont {L.}~\bibnamefont
  {Zhang}}, \bibinfo {author} {\bibfnamefont {D.}~\bibnamefont {Zhao}},
  \bibinfo {author} {\bibfnamefont {G.}~\bibnamefont {Jiang}}, \bibinfo
  {author} {\bibfnamefont {S.-C.}\ \bibnamefont {Zhang}}, \bibinfo {author}
  {\bibfnamefont {K.}~\bibnamefont {He}}, \bibinfo {author} {\bibfnamefont
  {X.}~\bibnamefont {Ma}}, \bibinfo {author} {\bibfnamefont {Q.-K.}\
  \bibnamefont {Xue}}, \ and\ \bibinfo {author} {\bibfnamefont
  {Y.}~\bibnamefont {Wang}},\ }\href {\doibase 10.1103/PhysRevLett.115.126801}
  {\bibfield  {journal} {\bibinfo  {journal} {Phys. Rev. Lett.}\ }\textbf
  {\bibinfo {volume} {115}},\ \bibinfo {pages} {126801} (\bibinfo {year}
  {2015})}\BibitemShut {NoStop}%
\bibitem [{\citenamefont {Kou}\ \emph {et~al.}(2014)\citenamefont {Kou},
  \citenamefont {Guo}, \citenamefont {Fan}, \citenamefont {Pan}, \citenamefont
  {Lang}, \citenamefont {Jiang}, \citenamefont {Shao}, \citenamefont {Nie},
  \citenamefont {Murata}, \citenamefont {Tang}, \citenamefont {Wang},
  \citenamefont {He}, \citenamefont {Lee}, \citenamefont {Lee},\ and\
  \citenamefont {Wang}}]{Kou2014}%
  \BibitemOpen
  \bibfield  {author} {\bibinfo {author} {\bibfnamefont {X.}~\bibnamefont
  {Kou}}, \bibinfo {author} {\bibfnamefont {S.-T.}\ \bibnamefont {Guo}},
  \bibinfo {author} {\bibfnamefont {Y.}~\bibnamefont {Fan}}, \bibinfo {author}
  {\bibfnamefont {L.}~\bibnamefont {Pan}}, \bibinfo {author} {\bibfnamefont
  {M.}~\bibnamefont {Lang}}, \bibinfo {author} {\bibfnamefont {Y.}~\bibnamefont
  {Jiang}}, \bibinfo {author} {\bibfnamefont {Q.}~\bibnamefont {Shao}},
  \bibinfo {author} {\bibfnamefont {T.}~\bibnamefont {Nie}}, \bibinfo {author}
  {\bibfnamefont {K.}~\bibnamefont {Murata}}, \bibinfo {author} {\bibfnamefont
  {J.}~\bibnamefont {Tang}}, \bibinfo {author} {\bibfnamefont {Y.}~\bibnamefont
  {Wang}}, \bibinfo {author} {\bibfnamefont {L.}~\bibnamefont {He}}, \bibinfo
  {author} {\bibfnamefont {T.-K.}\ \bibnamefont {Lee}}, \bibinfo {author}
  {\bibfnamefont {W.-L.}\ \bibnamefont {Lee}}, \ and\ \bibinfo {author}
  {\bibfnamefont {K.~L.}\ \bibnamefont {Wang}},\ }\href {\doibase
  10.1103/PhysRevLett.113.137201} {\bibfield  {journal} {\bibinfo  {journal}
  {Phys. Rev. Lett.}\ }\textbf {\bibinfo {volume} {113}},\ \bibinfo {pages}
  {137201} (\bibinfo {year} {2014})}\BibitemShut {NoStop}%
\bibitem [{\citenamefont {Bestwick}\ \emph {et~al.}(2015)\citenamefont
  {Bestwick}, \citenamefont {Fox}, \citenamefont {Kou}, \citenamefont {Pan},
  \citenamefont {Wang},\ and\ \citenamefont {Goldhaber-Gordon}}]{Bestwick2015}%
  \BibitemOpen
  \bibfield  {author} {\bibinfo {author} {\bibfnamefont {A.~J.}\ \bibnamefont
  {Bestwick}}, \bibinfo {author} {\bibfnamefont {E.~J.}\ \bibnamefont {Fox}},
  \bibinfo {author} {\bibfnamefont {X.}~\bibnamefont {Kou}}, \bibinfo {author}
  {\bibfnamefont {L.}~\bibnamefont {Pan}}, \bibinfo {author} {\bibfnamefont
  {K.~L.}\ \bibnamefont {Wang}}, \ and\ \bibinfo {author} {\bibfnamefont
  {D.}~\bibnamefont {Goldhaber-Gordon}},\ }\href {\doibase
  10.1103/PhysRevLett.114.187201} {\bibfield  {journal} {\bibinfo  {journal}
  {Phys. Rev. Lett.}\ }\textbf {\bibinfo {volume} {114}},\ \bibinfo {pages}
  {187201} (\bibinfo {year} {2015})}\BibitemShut {NoStop}%
\bibitem [{\citenamefont {Chang}\ \emph {et~al.}(2016)\citenamefont {Chang},
  \citenamefont {Zhao}, \citenamefont {Li}, \citenamefont {Jain}, \citenamefont
  {Liu}, \citenamefont {Moodera},\ and\ \citenamefont {Chan}}]{Chang2016Obse}%
  \BibitemOpen
  \bibfield  {author} {\bibinfo {author} {\bibfnamefont {C.-Z.}\ \bibnamefont
  {Chang}}, \bibinfo {author} {\bibfnamefont {W.}~\bibnamefont {Zhao}},
  \bibinfo {author} {\bibfnamefont {J.}~\bibnamefont {Li}}, \bibinfo {author}
  {\bibfnamefont {J.~K.}\ \bibnamefont {Jain}}, \bibinfo {author}
  {\bibfnamefont {C.}~\bibnamefont {Liu}}, \bibinfo {author} {\bibfnamefont
  {J.~S.}\ \bibnamefont {Moodera}}, \ and\ \bibinfo {author} {\bibfnamefont
  {M.~H.~W.}\ \bibnamefont {Chan}},\ }\href {\doibase
  10.1103/PhysRevLett.117.126802} {\bibfield  {journal} {\bibinfo  {journal}
  {Phys. Rev. Lett.}\ }\textbf {\bibinfo {volume} {117}},\ \bibinfo {pages}
  {126802} (\bibinfo {year} {2016})}\BibitemShut {NoStop}%
\bibitem [{\citenamefont {Yasuda}\ \emph {et~al.}(2017)\citenamefont {Yasuda},
  \citenamefont {Mogi}, \citenamefont {Yoshimi}, \citenamefont {Tsukazaki},
  \citenamefont {Takahashi}, \citenamefont {Kawasaki}, \citenamefont {Kagawa},\
  and\ \citenamefont {Tokura}}]{Yasuda2017Quantized}%
  \BibitemOpen
  \bibfield  {author} {\bibinfo {author} {\bibfnamefont {K.}~\bibnamefont
  {Yasuda}}, \bibinfo {author} {\bibfnamefont {M.}~\bibnamefont {Mogi}},
  \bibinfo {author} {\bibfnamefont {R.}~\bibnamefont {Yoshimi}}, \bibinfo
  {author} {\bibfnamefont {A.}~\bibnamefont {Tsukazaki}}, \bibinfo {author}
  {\bibfnamefont {K.~S.}\ \bibnamefont {Takahashi}}, \bibinfo {author}
  {\bibfnamefont {M.}~\bibnamefont {Kawasaki}}, \bibinfo {author}
  {\bibfnamefont {F.}~\bibnamefont {Kagawa}}, \ and\ \bibinfo {author}
  {\bibfnamefont {Y.}~\bibnamefont {Tokura}},\ }\href {\doibase
  10.1126/science.aan5991} {\bibfield  {journal} {\bibinfo  {journal}
  {Science}\ }\textbf {\bibinfo {volume} {358}},\ \bibinfo {pages} {1311}
  (\bibinfo {year} {2017})}\BibitemShut {NoStop}%
\bibitem [{\citenamefont {Sharpe}\ \emph {et~al.}(2019)\citenamefont {Sharpe},
  \citenamefont {Fox}, \citenamefont {Barnard}, \citenamefont {Finney},
  \citenamefont {Watanabe}, \citenamefont {Taniguchi}, \citenamefont
  {Kastner},\ and\ \citenamefont {Goldhaber-Gordon}}]{sharpe2019emergent}%
  \BibitemOpen
  \bibfield  {author} {\bibinfo {author} {\bibfnamefont {A.~L.}\ \bibnamefont
  {Sharpe}}, \bibinfo {author} {\bibfnamefont {E.~J.}\ \bibnamefont {Fox}},
  \bibinfo {author} {\bibfnamefont {A.~W.}\ \bibnamefont {Barnard}}, \bibinfo
  {author} {\bibfnamefont {J.}~\bibnamefont {Finney}}, \bibinfo {author}
  {\bibfnamefont {K.}~\bibnamefont {Watanabe}}, \bibinfo {author}
  {\bibfnamefont {T.}~\bibnamefont {Taniguchi}}, \bibinfo {author}
  {\bibfnamefont {M.}~\bibnamefont {Kastner}}, \ and\ \bibinfo {author}
  {\bibfnamefont {D.}~\bibnamefont {Goldhaber-Gordon}},\ }\href {\doibase
  10.1126/science.aaw3780} {\bibfield  {journal} {\bibinfo  {journal}
  {Science}\ }\textbf {\bibinfo {volume} {365}},\ \bibinfo {pages} {605}
  (\bibinfo {year} {2019})}\BibitemShut {NoStop}%
\bibitem [{\citenamefont {Serlin}\ \emph {et~al.}(2020)\citenamefont {Serlin},
  \citenamefont {Tschirhart}, \citenamefont {Polshyn}, \citenamefont {Zhang},
  \citenamefont {Zhu}, \citenamefont {Watanabe}, \citenamefont {Taniguchi},
  \citenamefont {Balents},\ and\ \citenamefont {Young}}]{serlin2020intrinsic}%
  \BibitemOpen
  \bibfield  {author} {\bibinfo {author} {\bibfnamefont {M.}~\bibnamefont
  {Serlin}}, \bibinfo {author} {\bibfnamefont {C.}~\bibnamefont {Tschirhart}},
  \bibinfo {author} {\bibfnamefont {H.}~\bibnamefont {Polshyn}}, \bibinfo
  {author} {\bibfnamefont {Y.}~\bibnamefont {Zhang}}, \bibinfo {author}
  {\bibfnamefont {J.}~\bibnamefont {Zhu}}, \bibinfo {author} {\bibfnamefont
  {K.}~\bibnamefont {Watanabe}}, \bibinfo {author} {\bibfnamefont
  {T.}~\bibnamefont {Taniguchi}}, \bibinfo {author} {\bibfnamefont
  {L.}~\bibnamefont {Balents}}, \ and\ \bibinfo {author} {\bibfnamefont
  {A.}~\bibnamefont {Young}},\ }\href {\doibase 10.1126/science.aay5533}
  {\bibfield  {journal} {\bibinfo  {journal} {Science}\ }\textbf {\bibinfo
  {volume} {367}},\ \bibinfo {pages} {900} (\bibinfo {year}
  {2020})}\BibitemShut {NoStop}%
\bibitem [{\citenamefont {Li}\ \emph {et~al.}(2009)\citenamefont {Li},
  \citenamefont {Chu}, \citenamefont {Jain},\ and\ \citenamefont
  {Shen}}]{Li2009}%
  \BibitemOpen
  \bibfield  {author} {\bibinfo {author} {\bibfnamefont {J.}~\bibnamefont
  {Li}}, \bibinfo {author} {\bibfnamefont {R.-L.}\ \bibnamefont {Chu}},
  \bibinfo {author} {\bibfnamefont {J.~K.}\ \bibnamefont {Jain}}, \ and\
  \bibinfo {author} {\bibfnamefont {S.-Q.}\ \bibnamefont {Shen}},\ }\href
  {\doibase 10.1103/PhysRevLett.102.136806} {\bibfield  {journal} {\bibinfo
  {journal} {Phys. Rev. Lett.}\ }\textbf {\bibinfo {volume} {102}},\ \bibinfo
  {pages} {136806} (\bibinfo {year} {2009})}\BibitemShut {NoStop}%
\bibitem [{\citenamefont {Groth}\ \emph {et~al.}(2009)\citenamefont {Groth},
  \citenamefont {Wimmer}, \citenamefont {Akhmerov}, \citenamefont
  {Tworzyd\l{}o},\ and\ \citenamefont {Beenakker}}]{Groth2009}%
  \BibitemOpen
  \bibfield  {author} {\bibinfo {author} {\bibfnamefont {C.~W.}\ \bibnamefont
  {Groth}}, \bibinfo {author} {\bibfnamefont {M.}~\bibnamefont {Wimmer}},
  \bibinfo {author} {\bibfnamefont {A.~R.}\ \bibnamefont {Akhmerov}}, \bibinfo
  {author} {\bibfnamefont {J.}~\bibnamefont {Tworzyd\l{}o}}, \ and\ \bibinfo
  {author} {\bibfnamefont {C.~W.~J.}\ \bibnamefont {Beenakker}},\ }\href
  {\doibase 10.1103/PhysRevLett.103.196805} {\bibfield  {journal} {\bibinfo
  {journal} {Phys. Rev. Lett.}\ }\textbf {\bibinfo {volume} {103}},\ \bibinfo
  {pages} {196805} (\bibinfo {year} {2009})}\BibitemShut {NoStop}%
\bibitem [{\citenamefont {Guo}\ \emph {et~al.}(2010)\citenamefont {Guo},
  \citenamefont {Rosenberg}, \citenamefont {Refael},\ and\ \citenamefont
  {Franz}}]{Guo10}%
  \BibitemOpen
  \bibfield  {author} {\bibinfo {author} {\bibfnamefont {H.-M.}\ \bibnamefont
  {Guo}}, \bibinfo {author} {\bibfnamefont {G.}~\bibnamefont {Rosenberg}},
  \bibinfo {author} {\bibfnamefont {G.}~\bibnamefont {Refael}}, \ and\ \bibinfo
  {author} {\bibfnamefont {M.}~\bibnamefont {Franz}},\ }\href {\doibase
  10.1103/PhysRevLett.105.216601} {\bibfield  {journal} {\bibinfo  {journal}
  {Phys. Rev. Lett.}\ }\textbf {\bibinfo {volume} {105}},\ \bibinfo {pages}
  {216601} (\bibinfo {year} {2010})}\BibitemShut {NoStop}%
\bibitem [{\citenamefont {Nomura}\ and\ \citenamefont
  {Nagaosa}(2011)}]{Namura2011}%
  \BibitemOpen
  \bibfield  {author} {\bibinfo {author} {\bibfnamefont {K.}~\bibnamefont
  {Nomura}}\ and\ \bibinfo {author} {\bibfnamefont {N.}~\bibnamefont
  {Nagaosa}},\ }\href {\doibase 10.1103/PhysRevLett.106.166802} {\bibfield
  {journal} {\bibinfo  {journal} {Phys. Rev. Lett.}\ }\textbf {\bibinfo
  {volume} {106}},\ \bibinfo {pages} {166802} (\bibinfo {year}
  {2011})}\BibitemShut {NoStop}%
\bibitem [{\citenamefont {Lu}\ \emph {et~al.}(2011)\citenamefont {Lu},
  \citenamefont {Shi},\ and\ \citenamefont {Shen}}]{LuHZ2011}%
  \BibitemOpen
  \bibfield  {author} {\bibinfo {author} {\bibfnamefont {H.-Z.}\ \bibnamefont
  {Lu}}, \bibinfo {author} {\bibfnamefont {J.}~\bibnamefont {Shi}}, \ and\
  \bibinfo {author} {\bibfnamefont {S.-Q.}\ \bibnamefont {Shen}},\ }\href
  {\doibase 10.1103/PhysRevLett.107.076801} {\bibfield  {journal} {\bibinfo
  {journal} {Phys. Rev. Lett.}\ }\textbf {\bibinfo {volume} {107}},\ \bibinfo
  {pages} {076801} (\bibinfo {year} {2011})}\BibitemShut {NoStop}%
\bibitem [{\citenamefont {Qiao}\ \emph {et~al.}(2016)\citenamefont {Qiao},
  \citenamefont {Han}, \citenamefont {Zhang}, \citenamefont {Wang},
  \citenamefont {Deng}, \citenamefont {Jiang}, \citenamefont {Yang},
  \citenamefont {Wang},\ and\ \citenamefont {Niu}}]{QiaoZH2016}%
  \BibitemOpen
  \bibfield  {author} {\bibinfo {author} {\bibfnamefont {Z.}~\bibnamefont
  {Qiao}}, \bibinfo {author} {\bibfnamefont {Y.}~\bibnamefont {Han}}, \bibinfo
  {author} {\bibfnamefont {L.}~\bibnamefont {Zhang}}, \bibinfo {author}
  {\bibfnamefont {K.}~\bibnamefont {Wang}}, \bibinfo {author} {\bibfnamefont
  {X.}~\bibnamefont {Deng}}, \bibinfo {author} {\bibfnamefont {H.}~\bibnamefont
  {Jiang}}, \bibinfo {author} {\bibfnamefont {S.~A.}\ \bibnamefont {Yang}},
  \bibinfo {author} {\bibfnamefont {J.}~\bibnamefont {Wang}}, \ and\ \bibinfo
  {author} {\bibfnamefont {Q.}~\bibnamefont {Niu}},\ }\href {\doibase
  10.1103/PhysRevLett.117.056802} {\bibfield  {journal} {\bibinfo  {journal}
  {Phys. Rev. Lett.}\ }\textbf {\bibinfo {volume} {117}},\ \bibinfo {pages}
  {056802} (\bibinfo {year} {2016})}\BibitemShut {NoStop}%
\bibitem [{\citenamefont {Chen}\ \emph
  {et~al.}(2019{\natexlab{a}})\citenamefont {Chen}, \citenamefont {Liu},\ and\
  \citenamefont {Xie}}]{ChenCZ2019}%
  \BibitemOpen
  \bibfield  {author} {\bibinfo {author} {\bibfnamefont {C.-Z.}\ \bibnamefont
  {Chen}}, \bibinfo {author} {\bibfnamefont {H.}~\bibnamefont {Liu}}, \ and\
  \bibinfo {author} {\bibfnamefont {X.~C.}\ \bibnamefont {Xie}},\ }\href
  {\doibase 10.1103/PhysRevLett.122.026601} {\bibfield  {journal} {\bibinfo
  {journal} {Phys. Rev. Lett.}\ }\textbf {\bibinfo {volume} {122}},\ \bibinfo
  {pages} {026601} (\bibinfo {year} {2019}{\natexlab{a}})}\BibitemShut
  {NoStop}%
\bibitem [{\citenamefont {Haim}\ \emph {et~al.}(2019)\citenamefont {Haim},
  \citenamefont {Ilan},\ and\ \citenamefont {Alicea}}]{Haim2019}%
  \BibitemOpen
  \bibfield  {author} {\bibinfo {author} {\bibfnamefont {A.}~\bibnamefont
  {Haim}}, \bibinfo {author} {\bibfnamefont {R.}~\bibnamefont {Ilan}}, \ and\
  \bibinfo {author} {\bibfnamefont {J.}~\bibnamefont {Alicea}},\ }\href
  {\doibase 10.1103/PhysRevLett.123.046801} {\bibfield  {journal} {\bibinfo
  {journal} {Phys. Rev. Lett.}\ }\textbf {\bibinfo {volume} {123}},\ \bibinfo
  {pages} {046801} (\bibinfo {year} {2019})}\BibitemShut {NoStop}%
\bibitem [{\citenamefont {Xing}\ \emph {et~al.}(2018)\citenamefont {Xing},
  \citenamefont {Xu}, \citenamefont {Cheung}, \citenamefont {Sun},
  \citenamefont {Wang},\ and\ \citenamefont {Yao}}]{xing2018influence}%
  \BibitemOpen
  \bibfield  {author} {\bibinfo {author} {\bibfnamefont {Y.}~\bibnamefont
  {Xing}}, \bibinfo {author} {\bibfnamefont {F.}~\bibnamefont {Xu}}, \bibinfo
  {author} {\bibfnamefont {K.~T.}\ \bibnamefont {Cheung}}, \bibinfo {author}
  {\bibfnamefont {Q.-f.}\ \bibnamefont {Sun}}, \bibinfo {author} {\bibfnamefont
  {J.}~\bibnamefont {Wang}}, \ and\ \bibinfo {author} {\bibfnamefont
  {Y.}~\bibnamefont {Yao}},\ }\href {\doibase 10.1088/1367-2630/aab4e8}
  {\bibfield  {journal} {\bibinfo  {journal} {New J. Phys.}\ }\textbf {\bibinfo
  {volume} {20}},\ \bibinfo {pages} {043011} (\bibinfo {year}
  {2018})}\BibitemShut {NoStop}%
\bibitem [{\citenamefont {Wang}\ \emph {et~al.}(2014)\citenamefont {Wang},
  \citenamefont {Lian},\ and\ \citenamefont {Zhang}}]{WangJ2014Uni}%
  \BibitemOpen
  \bibfield  {author} {\bibinfo {author} {\bibfnamefont {J.}~\bibnamefont
  {Wang}}, \bibinfo {author} {\bibfnamefont {B.}~\bibnamefont {Lian}}, \ and\
  \bibinfo {author} {\bibfnamefont {S.-C.}\ \bibnamefont {Zhang}},\ }\href
  {\doibase 10.1103/PhysRevB.89.085106} {\bibfield  {journal} {\bibinfo
  {journal} {Phys. Rev. B}\ }\textbf {\bibinfo {volume} {89}},\ \bibinfo
  {pages} {085106} (\bibinfo {year} {2014})}\BibitemShut {NoStop}%
\bibitem [{\citenamefont {Keser}\ \emph {et~al.}(2019)\citenamefont {Keser},
  \citenamefont {Raimondi},\ and\ \citenamefont {Culcer}}]{Keser2019}%
  \BibitemOpen
  \bibfield  {author} {\bibinfo {author} {\bibfnamefont {A.~C.}\ \bibnamefont
  {Keser}}, \bibinfo {author} {\bibfnamefont {R.}~\bibnamefont {Raimondi}}, \
  and\ \bibinfo {author} {\bibfnamefont {D.}~\bibnamefont {Culcer}},\ }\href
  {\doibase 10.1103/PhysRevLett.123.126603} {\bibfield  {journal} {\bibinfo
  {journal} {Phys. Rev. Lett.}\ }\textbf {\bibinfo {volume} {123}},\ \bibinfo
  {pages} {126603} (\bibinfo {year} {2019})}\BibitemShut {NoStop}%
\bibitem [{\citenamefont {Wang}\ \emph {et~al.}(2018)\citenamefont {Wang},
  \citenamefont {Ou}, \citenamefont {Liu}, \citenamefont {Wang}, \citenamefont
  {He}, \citenamefont {Xue},\ and\ \citenamefont {Wu}}]{wang2018direct}%
  \BibitemOpen
  \bibfield  {author} {\bibinfo {author} {\bibfnamefont {W.}~\bibnamefont
  {Wang}}, \bibinfo {author} {\bibfnamefont {Y.}~\bibnamefont {Ou}}, \bibinfo
  {author} {\bibfnamefont {C.}~\bibnamefont {Liu}}, \bibinfo {author}
  {\bibfnamefont {Y.}~\bibnamefont {Wang}}, \bibinfo {author} {\bibfnamefont
  {K.}~\bibnamefont {He}}, \bibinfo {author} {\bibfnamefont {Q.-K.}\
  \bibnamefont {Xue}}, \ and\ \bibinfo {author} {\bibfnamefont
  {W.}~\bibnamefont {Wu}},\ }\href {\doibase 10.1038/s41567-018-0149-1}
  {\bibfield  {journal} {\bibinfo  {journal} {Nat. Phys.}\ }\textbf {\bibinfo
  {volume} {14}},\ \bibinfo {pages} {791} (\bibinfo {year} {2018})}\BibitemShut
  {NoStop}%
\bibitem [{\citenamefont {Lee}\ \emph {et~al.}(2015)\citenamefont {Lee},
  \citenamefont {Kim}, \citenamefont {Lee}, \citenamefont {Billinge},
  \citenamefont {Zhong}, \citenamefont {Schneeloch}, \citenamefont {Liu},
  \citenamefont {Valla}, \citenamefont {Tranquada}, \citenamefont {Gu} \emph
  {et~al.}}]{lee2015imaging}%
  \BibitemOpen
  \bibfield  {author} {\bibinfo {author} {\bibfnamefont {I.}~\bibnamefont
  {Lee}}, \bibinfo {author} {\bibfnamefont {C.~K.}\ \bibnamefont {Kim}},
  \bibinfo {author} {\bibfnamefont {J.}~\bibnamefont {Lee}}, \bibinfo {author}
  {\bibfnamefont {S.~J.}\ \bibnamefont {Billinge}}, \bibinfo {author}
  {\bibfnamefont {R.}~\bibnamefont {Zhong}}, \bibinfo {author} {\bibfnamefont
  {J.~A.}\ \bibnamefont {Schneeloch}}, \bibinfo {author} {\bibfnamefont
  {T.}~\bibnamefont {Liu}}, \bibinfo {author} {\bibfnamefont {T.}~\bibnamefont
  {Valla}}, \bibinfo {author} {\bibfnamefont {J.~M.}\ \bibnamefont
  {Tranquada}}, \bibinfo {author} {\bibfnamefont {G.}~\bibnamefont {Gu}},
  \emph {et~al.},\ }\href {\doibase 10.1073/pnas.1424322112} {\bibfield
  {journal} {\bibinfo  {journal} {PNAS}\ }\textbf {\bibinfo {volume} {112}},\
  \bibinfo {pages} {1316} (\bibinfo {year} {2015})}\BibitemShut {NoStop}%
\bibitem [{\citenamefont {Lachman}\ \emph {et~al.}(2015)\citenamefont
  {Lachman}, \citenamefont {Young}, \citenamefont {Richardella}, \citenamefont
  {Cuppens}, \citenamefont {Naren}, \citenamefont {Anahory}, \citenamefont
  {Meltzer}, \citenamefont {Kandala}, \citenamefont {Kempinger}, \citenamefont
  {Myasoedov} \emph {et~al.}}]{lachman2015vis}%
  \BibitemOpen
  \bibfield  {author} {\bibinfo {author} {\bibfnamefont {E.~O.}\ \bibnamefont
  {Lachman}}, \bibinfo {author} {\bibfnamefont {A.~F.}\ \bibnamefont {Young}},
  \bibinfo {author} {\bibfnamefont {A.}~\bibnamefont {Richardella}}, \bibinfo
  {author} {\bibfnamefont {J.}~\bibnamefont {Cuppens}}, \bibinfo {author}
  {\bibfnamefont {H.}~\bibnamefont {Naren}}, \bibinfo {author} {\bibfnamefont
  {Y.}~\bibnamefont {Anahory}}, \bibinfo {author} {\bibfnamefont {A.~Y.}\
  \bibnamefont {Meltzer}}, \bibinfo {author} {\bibfnamefont {A.}~\bibnamefont
  {Kandala}}, \bibinfo {author} {\bibfnamefont {S.}~\bibnamefont {Kempinger}},
  \bibinfo {author} {\bibfnamefont {Y.}~\bibnamefont {Myasoedov}},  \emph
  {et~al.},\ }\href {\doibase 10.1126/sciadv.1500740} {\bibfield  {journal}
  {\bibinfo  {journal} {Sci. Adv.}\ }\textbf {\bibinfo {volume} {1}},\ \bibinfo
  {pages} {e1500740} (\bibinfo {year} {2015})}\BibitemShut {NoStop}%
\bibitem [{\citenamefont {Kou}\ \emph {et~al.}(2015)\citenamefont {Kou},
  \citenamefont {Pan}, \citenamefont {Wang}, \citenamefont {Fan}, \citenamefont
  {Choi}, \citenamefont {Lee}, \citenamefont {Nie}, \citenamefont {Murata},
  \citenamefont {Shao}, \citenamefont {Zhang} \emph {et~al.}}]{kou2015metal}%
  \BibitemOpen
  \bibfield  {author} {\bibinfo {author} {\bibfnamefont {X.}~\bibnamefont
  {Kou}}, \bibinfo {author} {\bibfnamefont {L.}~\bibnamefont {Pan}}, \bibinfo
  {author} {\bibfnamefont {J.}~\bibnamefont {Wang}}, \bibinfo {author}
  {\bibfnamefont {Y.}~\bibnamefont {Fan}}, \bibinfo {author} {\bibfnamefont
  {E.~S.}\ \bibnamefont {Choi}}, \bibinfo {author} {\bibfnamefont {W.-L.}\
  \bibnamefont {Lee}}, \bibinfo {author} {\bibfnamefont {T.}~\bibnamefont
  {Nie}}, \bibinfo {author} {\bibfnamefont {K.}~\bibnamefont {Murata}},
  \bibinfo {author} {\bibfnamefont {Q.}~\bibnamefont {Shao}}, \bibinfo {author}
  {\bibfnamefont {S.-C.}\ \bibnamefont {Zhang}},  \emph {et~al.},\ }\href
  {\doibase doi.org/10.1038/ncomms9474} {\bibfield  {journal} {\bibinfo
  {journal} {Nat. Comm.}\ }\textbf {\bibinfo {volume} {6}},\ \bibinfo {pages}
  {1} (\bibinfo {year} {2015})}\BibitemShut {NoStop}%
\bibitem [{\citenamefont {Yuan}\ \emph {et~al.}(2020)\citenamefont {Yuan},
  \citenamefont {Wang}, \citenamefont {Li}, \citenamefont {Li}, \citenamefont
  {Ji}, \citenamefont {Hao}, \citenamefont {Wu}, \citenamefont {He},
  \citenamefont {Wang}, \citenamefont {Xu}, \citenamefont {Duan}, \citenamefont
  {Li},\ and\ \citenamefont {Xue}}]{YuanYH2020}%
  \BibitemOpen
  \bibfield  {author} {\bibinfo {author} {\bibfnamefont {Y.}~\bibnamefont
  {Yuan}}, \bibinfo {author} {\bibfnamefont {X.}~\bibnamefont {Wang}}, \bibinfo
  {author} {\bibfnamefont {H.}~\bibnamefont {Li}}, \bibinfo {author}
  {\bibfnamefont {J.}~\bibnamefont {Li}}, \bibinfo {author} {\bibfnamefont
  {Y.}~\bibnamefont {Ji}}, \bibinfo {author} {\bibfnamefont {Z.}~\bibnamefont
  {Hao}}, \bibinfo {author} {\bibfnamefont {Y.}~\bibnamefont {Wu}}, \bibinfo
  {author} {\bibfnamefont {K.}~\bibnamefont {He}}, \bibinfo {author}
  {\bibfnamefont {Y.}~\bibnamefont {Wang}}, \bibinfo {author} {\bibfnamefont
  {Y.}~\bibnamefont {Xu}}, \bibinfo {author} {\bibfnamefont {W.}~\bibnamefont
  {Duan}}, \bibinfo {author} {\bibfnamefont {W.}~\bibnamefont {Li}}, \ and\
  \bibinfo {author} {\bibfnamefont {Q.-K.}\ \bibnamefont {Xue}},\ }\href
  {\doibase 10.1021/acs.nanolett.0c00031} {\bibfield  {journal} {\bibinfo
  {journal} {Nano Lett.}\ }\textbf {\bibinfo {volume} {20}},\ \bibinfo {pages}
  {3271} (\bibinfo {year} {2020})}\BibitemShut {NoStop}%
\bibitem [{\citenamefont {Chen}\ \emph {et~al.}(2015)\citenamefont {Chen},
  \citenamefont {Teague}, \citenamefont {He}, \citenamefont {Kou},
  \citenamefont {Lang}, \citenamefont {Fan}, \citenamefont {Woodward},
  \citenamefont {Wang},\ and\ \citenamefont {Yeh}}]{chen2015magnetism}%
  \BibitemOpen
  \bibfield  {author} {\bibinfo {author} {\bibfnamefont {C.}~\bibnamefont
  {Chen}}, \bibinfo {author} {\bibfnamefont {M.}~\bibnamefont {Teague}},
  \bibinfo {author} {\bibfnamefont {L.}~\bibnamefont {He}}, \bibinfo {author}
  {\bibfnamefont {X.}~\bibnamefont {Kou}}, \bibinfo {author} {\bibfnamefont
  {M.}~\bibnamefont {Lang}}, \bibinfo {author} {\bibfnamefont {W.}~\bibnamefont
  {Fan}}, \bibinfo {author} {\bibfnamefont {N.}~\bibnamefont {Woodward}},
  \bibinfo {author} {\bibfnamefont {K.}~\bibnamefont {Wang}}, \ and\ \bibinfo
  {author} {\bibfnamefont {N.}~\bibnamefont {Yeh}},\ }\href {\doibase
  10.1088/1367-2630/17/11/113042} {\bibfield  {journal} {\bibinfo  {journal}
  {New J. Phys.}\ }\textbf {\bibinfo {volume} {17}},\ \bibinfo {pages} {113042}
  (\bibinfo {year} {2015})}\BibitemShut {NoStop}%
\bibitem [{\citenamefont {Liao}\ \emph {et~al.}(2015)\citenamefont {Liao},
  \citenamefont {Ou}, \citenamefont {Feng}, \citenamefont {Yang}, \citenamefont
  {Lin}, \citenamefont {Yang}, \citenamefont {Wu}, \citenamefont {He},
  \citenamefont {Ma}, \citenamefont {Xue},\ and\ \citenamefont
  {Li}}]{LiaoJ2015}%
  \BibitemOpen
  \bibfield  {author} {\bibinfo {author} {\bibfnamefont {J.}~\bibnamefont
  {Liao}}, \bibinfo {author} {\bibfnamefont {Y.}~\bibnamefont {Ou}}, \bibinfo
  {author} {\bibfnamefont {X.}~\bibnamefont {Feng}}, \bibinfo {author}
  {\bibfnamefont {S.}~\bibnamefont {Yang}}, \bibinfo {author} {\bibfnamefont
  {C.}~\bibnamefont {Lin}}, \bibinfo {author} {\bibfnamefont {W.}~\bibnamefont
  {Yang}}, \bibinfo {author} {\bibfnamefont {K.}~\bibnamefont {Wu}}, \bibinfo
  {author} {\bibfnamefont {K.}~\bibnamefont {He}}, \bibinfo {author}
  {\bibfnamefont {X.}~\bibnamefont {Ma}}, \bibinfo {author} {\bibfnamefont
  {Q.-K.}\ \bibnamefont {Xue}}, \ and\ \bibinfo {author} {\bibfnamefont
  {Y.}~\bibnamefont {Li}},\ }\href {\doibase 10.1103/PhysRevLett.114.216601}
  {\bibfield  {journal} {\bibinfo  {journal} {Phys. Rev. Lett.}\ }\textbf
  {\bibinfo {volume} {114}},\ \bibinfo {pages} {216601} (\bibinfo {year}
  {2015})}\BibitemShut {NoStop}%
\bibitem [{\citenamefont {Xing}\ \emph {et~al.}(2011)\citenamefont {Xing},
  \citenamefont {Zhang},\ and\ \citenamefont {Wang}}]{Yanxia11}%
  \BibitemOpen
  \bibfield  {author} {\bibinfo {author} {\bibfnamefont {Y.}~\bibnamefont
  {Xing}}, \bibinfo {author} {\bibfnamefont {L.}~\bibnamefont {Zhang}}, \ and\
  \bibinfo {author} {\bibfnamefont {J.}~\bibnamefont {Wang}},\ }\href {\doibase
  10.1103/PhysRevB.84.035110} {\bibfield  {journal} {\bibinfo  {journal} {Phys.
  Rev. B}\ }\textbf {\bibinfo {volume} {84}},\ \bibinfo {pages} {035110}
  (\bibinfo {year} {2011})}\BibitemShut {NoStop}%
\bibitem [{\citenamefont {Prodan}(2011)}]{Prodan11}%
  \BibitemOpen
  \bibfield  {author} {\bibinfo {author} {\bibfnamefont {E.}~\bibnamefont
  {Prodan}},\ }\href {\doibase 10.1103/PhysRevB.83.195119} {\bibfield
  {journal} {\bibinfo  {journal} {Phys. Rev. B}\ }\textbf {\bibinfo {volume}
  {83}},\ \bibinfo {pages} {195119} (\bibinfo {year} {2011})}\BibitemShut
  {NoStop}%
\bibitem [{\citenamefont {Yamakage}\ \emph {et~al.}(2011)\citenamefont
  {Yamakage}, \citenamefont {Nomura}, \citenamefont {Imura},\ and\
  \citenamefont {Kuramoto}}]{Yamakage11}%
  \BibitemOpen
  \bibfield  {author} {\bibinfo {author} {\bibfnamefont {A.}~\bibnamefont
  {Yamakage}}, \bibinfo {author} {\bibfnamefont {K.}~\bibnamefont {Nomura}},
  \bibinfo {author} {\bibfnamefont {K.-I.}\ \bibnamefont {Imura}}, \ and\
  \bibinfo {author} {\bibfnamefont {Y.}~\bibnamefont {Kuramoto}},\ }\href
  {\doibase 10.1143/JPSJ.80.053703} {\bibfield  {journal} {\bibinfo  {journal}
  {Journal of the Physical Society of Japan}\ }\textbf {\bibinfo {volume}
  {80}},\ \bibinfo {pages} {053703} (\bibinfo {year} {2011})},\ \Eprint
  {http://arxiv.org/abs/https://doi.org/10.1143/JPSJ.80.053703}
  {https://doi.org/10.1143/JPSJ.80.053703} \BibitemShut {NoStop}%
\bibitem [{\citenamefont {Zhang}\ \emph {et~al.}(2012)\citenamefont {Zhang},
  \citenamefont {Chu}, \citenamefont {Zhang},\ and\ \citenamefont
  {Shen}}]{Yan-Yang12}%
  \BibitemOpen
  \bibfield  {author} {\bibinfo {author} {\bibfnamefont {Y.-Y.}\ \bibnamefont
  {Zhang}}, \bibinfo {author} {\bibfnamefont {R.-L.}\ \bibnamefont {Chu}},
  \bibinfo {author} {\bibfnamefont {F.-C.}\ \bibnamefont {Zhang}}, \ and\
  \bibinfo {author} {\bibfnamefont {S.-Q.}\ \bibnamefont {Shen}},\ }\href
  {\doibase 10.1103/PhysRevB.85.035107} {\bibfield  {journal} {\bibinfo
  {journal} {Phys. Rev. B}\ }\textbf {\bibinfo {volume} {85}},\ \bibinfo
  {pages} {035107} (\bibinfo {year} {2012})}\BibitemShut {NoStop}%
\bibitem [{\citenamefont {Xu}\ \emph {et~al.}(2012)\citenamefont {Xu},
  \citenamefont {Qi}, \citenamefont {Liu}, \citenamefont {Sacksteder},
  \citenamefont {Xie},\ and\ \citenamefont {Jiang}}]{Dongwei12}%
  \BibitemOpen
  \bibfield  {author} {\bibinfo {author} {\bibfnamefont {D.}~\bibnamefont
  {Xu}}, \bibinfo {author} {\bibfnamefont {J.}~\bibnamefont {Qi}}, \bibinfo
  {author} {\bibfnamefont {J.}~\bibnamefont {Liu}}, \bibinfo {author}
  {\bibfnamefont {V.}~\bibnamefont {Sacksteder}}, \bibinfo {author}
  {\bibfnamefont {X.~C.}\ \bibnamefont {Xie}}, \ and\ \bibinfo {author}
  {\bibfnamefont {H.}~\bibnamefont {Jiang}},\ }\href {\doibase
  10.1103/PhysRevB.85.195140} {\bibfield  {journal} {\bibinfo  {journal} {Phys.
  Rev. B}\ }\textbf {\bibinfo {volume} {85}},\ \bibinfo {pages} {195140}
  (\bibinfo {year} {2012})}\BibitemShut {NoStop}%
\bibitem [{\citenamefont {Okugawa}\ \emph {et~al.}(2020)\citenamefont
  {Okugawa}, \citenamefont {Tang}, \citenamefont {Rubio},\ and\ \citenamefont
  {Kennes}}]{Okugawa20}%
  \BibitemOpen
  \bibfield  {author} {\bibinfo {author} {\bibfnamefont {T.}~\bibnamefont
  {Okugawa}}, \bibinfo {author} {\bibfnamefont {P.}~\bibnamefont {Tang}},
  \bibinfo {author} {\bibfnamefont {A.}~\bibnamefont {Rubio}}, \ and\ \bibinfo
  {author} {\bibfnamefont {D.~M.}\ \bibnamefont {Kennes}},\ }\href {\doibase
  10.1103/PhysRevB.102.201405} {\bibfield  {journal} {\bibinfo  {journal}
  {Phys. Rev. B}\ }\textbf {\bibinfo {volume} {102}},\ \bibinfo {pages}
  {201405} (\bibinfo {year} {2020})}\BibitemShut {NoStop}%
\bibitem [{\citenamefont {Zhang}\ \emph {et~al.}(2021)\citenamefont {Zhang},
  \citenamefont {Chen}, \citenamefont {Wu}, \citenamefont {Jiang},
  \citenamefont {Liu}, \citenamefont {Sun},\ and\ \citenamefont
  {Xie}}]{Zhi-Qiang21}%
  \BibitemOpen
  \bibfield  {author} {\bibinfo {author} {\bibfnamefont {Z.-Q.}\ \bibnamefont
  {Zhang}}, \bibinfo {author} {\bibfnamefont {C.-Z.}\ \bibnamefont {Chen}},
  \bibinfo {author} {\bibfnamefont {Y.}~\bibnamefont {Wu}}, \bibinfo {author}
  {\bibfnamefont {H.}~\bibnamefont {Jiang}}, \bibinfo {author} {\bibfnamefont
  {J.}~\bibnamefont {Liu}}, \bibinfo {author} {\bibfnamefont {Q.-f.}\
  \bibnamefont {Sun}}, \ and\ \bibinfo {author} {\bibfnamefont {X.~C.}\
  \bibnamefont {Xie}},\ }\href {\doibase 10.1103/PhysRevB.103.075434}
  {\bibfield  {journal} {\bibinfo  {journal} {Phys. Rev. B}\ }\textbf {\bibinfo
  {volume} {103}},\ \bibinfo {pages} {075434} (\bibinfo {year}
  {2021})}\BibitemShut {NoStop}%
\bibitem [{\citenamefont {Song}\ \emph {et~al.}(2012)\citenamefont {Song},
  \citenamefont {Liu}, \citenamefont {Jiang}, \citenamefont {Sun},\ and\
  \citenamefont {Xie}}]{Song12}%
  \BibitemOpen
  \bibfield  {author} {\bibinfo {author} {\bibfnamefont {J.}~\bibnamefont
  {Song}}, \bibinfo {author} {\bibfnamefont {H.}~\bibnamefont {Liu}}, \bibinfo
  {author} {\bibfnamefont {H.}~\bibnamefont {Jiang}}, \bibinfo {author}
  {\bibfnamefont {Q.-f.}\ \bibnamefont {Sun}}, \ and\ \bibinfo {author}
  {\bibfnamefont {X.~C.}\ \bibnamefont {Xie}},\ }\href {\doibase
  10.1103/PhysRevB.85.195125} {\bibfield  {journal} {\bibinfo  {journal} {Phys.
  Rev. B}\ }\textbf {\bibinfo {volume} {85}},\ \bibinfo {pages} {195125}
  (\bibinfo {year} {2012})}\BibitemShut {NoStop}%
\bibitem [{\citenamefont {Aubry}\ and\ \citenamefont
  {Andr{\'e}}()}]{aubry1980analyticity}%
  \BibitemOpen
  \bibfield  {author} {\bibinfo {author} {\bibfnamefont {S.}~\bibnamefont
  {Aubry}}\ and\ \bibinfo {author} {\bibfnamefont {G.}~\bibnamefont
  {Andr{\'e}}},\ }\href@noop {} {\bibfield  {journal} {\bibinfo  {journal}
  {Ann. Israel Phys. Soc}\ }\textbf {\bibinfo {volume} {3}},\ \bibinfo {pages}
  {18}}\BibitemShut {NoStop}%
\bibitem [{\citenamefont {Girschik}\ \emph {et~al.}(2013)\citenamefont
  {Girschik}, \citenamefont {Libisch},\ and\ \citenamefont
  {Rotter}}]{Girschik13}%
  \BibitemOpen
  \bibfield  {author} {\bibinfo {author} {\bibfnamefont {A.}~\bibnamefont
  {Girschik}}, \bibinfo {author} {\bibfnamefont {F.}~\bibnamefont {Libisch}}, \
  and\ \bibinfo {author} {\bibfnamefont {S.}~\bibnamefont {Rotter}},\ }\href
  {\doibase 10.1103/PhysRevB.88.014201} {\bibfield  {journal} {\bibinfo
  {journal} {Phys. Rev. B}\ }\textbf {\bibinfo {volume} {88}},\ \bibinfo
  {pages} {014201} (\bibinfo {year} {2013})}\BibitemShut {NoStop}%
\bibitem [{\citenamefont {Fu}\ \emph {et~al.}(2021)\citenamefont {Fu},
  \citenamefont {Wilson},\ and\ \citenamefont {Pixley}}]{Fu21}%
  \BibitemOpen
  \bibfield  {author} {\bibinfo {author} {\bibfnamefont {Y.}~\bibnamefont
  {Fu}}, \bibinfo {author} {\bibfnamefont {J.~H.}\ \bibnamefont {Wilson}}, \
  and\ \bibinfo {author} {\bibfnamefont {J.~H.}\ \bibnamefont {Pixley}},\
  }\href {\doibase 10.1103/PhysRevB.104.L041106} {\bibfield  {journal}
  {\bibinfo  {journal} {Phys. Rev. B}\ }\textbf {\bibinfo {volume} {104}},\
  \bibinfo {pages} {L041106} (\bibinfo {year} {2021})}\BibitemShut {NoStop}%
\bibitem [{\citenamefont {Chen}\ \emph
  {et~al.}(2019{\natexlab{b}})\citenamefont {Chen}, \citenamefont {Xu},\ and\
  \citenamefont {Zhou}}]{Dong-Hui19}%
  \BibitemOpen
  \bibfield  {author} {\bibinfo {author} {\bibfnamefont {R.}~\bibnamefont
  {Chen}}, \bibinfo {author} {\bibfnamefont {D.-H.}\ \bibnamefont {Xu}}, \ and\
  \bibinfo {author} {\bibfnamefont {B.}~\bibnamefont {Zhou}},\ }\href {\doibase
  10.1103/PhysRevB.100.115311} {\bibfield  {journal} {\bibinfo  {journal}
  {Phys. Rev. B}\ }\textbf {\bibinfo {volume} {100}},\ \bibinfo {pages}
  {115311} (\bibinfo {year} {2019}{\natexlab{b}})}\BibitemShut {NoStop}%
\bibitem [{\citenamefont {Chen}\ \emph {et~al.}(2018)\citenamefont {Chen},
  \citenamefont {Chen}, \citenamefont {Sun}, \citenamefont {Zhou},\ and\
  \citenamefont {Xu}}]{Rui18}%
  \BibitemOpen
  \bibfield  {author} {\bibinfo {author} {\bibfnamefont {R.}~\bibnamefont
  {Chen}}, \bibinfo {author} {\bibfnamefont {C.-Z.}\ \bibnamefont {Chen}},
  \bibinfo {author} {\bibfnamefont {J.-H.}\ \bibnamefont {Sun}}, \bibinfo
  {author} {\bibfnamefont {B.}~\bibnamefont {Zhou}}, \ and\ \bibinfo {author}
  {\bibfnamefont {D.-H.}\ \bibnamefont {Xu}},\ }\href {\doibase
  10.1103/PhysRevB.97.235109} {\bibfield  {journal} {\bibinfo  {journal} {Phys.
  Rev. B}\ }\textbf {\bibinfo {volume} {97}},\ \bibinfo {pages} {235109}
  (\bibinfo {year} {2018})}\BibitemShut {NoStop}%
\bibitem [{\citenamefont {Biddle}\ \emph {et~al.}(2011)\citenamefont {Biddle},
  \citenamefont {Priour}, \citenamefont {Wang},\ and\ \citenamefont
  {Das~Sarma}}]{Biddle11}%
  \BibitemOpen
  \bibfield  {author} {\bibinfo {author} {\bibfnamefont {J.}~\bibnamefont
  {Biddle}}, \bibinfo {author} {\bibfnamefont {D.~J.}\ \bibnamefont {Priour}},
  \bibinfo {author} {\bibfnamefont {B.}~\bibnamefont {Wang}}, \ and\ \bibinfo
  {author} {\bibfnamefont {S.}~\bibnamefont {Das~Sarma}},\ }\href {\doibase
  10.1103/PhysRevB.83.075105} {\bibfield  {journal} {\bibinfo  {journal} {Phys.
  Rev. B}\ }\textbf {\bibinfo {volume} {83}},\ \bibinfo {pages} {075105}
  (\bibinfo {year} {2011})}\BibitemShut {NoStop}%
\bibitem [{\citenamefont {Modak}\ and\ \citenamefont
  {Mukerjee}(2015)}]{Modak15}%
  \BibitemOpen
  \bibfield  {author} {\bibinfo {author} {\bibfnamefont {R.}~\bibnamefont
  {Modak}}\ and\ \bibinfo {author} {\bibfnamefont {S.}~\bibnamefont
  {Mukerjee}},\ }\href {\doibase 10.1103/PhysRevLett.115.230401} {\bibfield
  {journal} {\bibinfo  {journal} {Phys. Rev. Lett.}\ }\textbf {\bibinfo
  {volume} {115}},\ \bibinfo {pages} {230401} (\bibinfo {year}
  {2015})}\BibitemShut {NoStop}%
\bibitem [{\citenamefont {Modak}\ and\ \citenamefont {Nag}(2020)}]{Modak20}%
  \BibitemOpen
  \bibfield  {author} {\bibinfo {author} {\bibfnamefont {R.}~\bibnamefont
  {Modak}}\ and\ \bibinfo {author} {\bibfnamefont {T.}~\bibnamefont {Nag}},\
  }\href {\doibase 10.1103/PhysRevResearch.2.012074} {\bibfield  {journal}
  {\bibinfo  {journal} {Phys. Rev. Research}\ }\textbf {\bibinfo {volume}
  {2}},\ \bibinfo {pages} {012074} (\bibinfo {year} {2020})}\BibitemShut
  {NoStop}%
\bibitem [{\citenamefont {Deng}\ \emph {et~al.}(2019)\citenamefont {Deng},
  \citenamefont {Ray}, \citenamefont {Sinha}, \citenamefont {Shlyapnikov},\
  and\ \citenamefont {Santos}}]{Deng19}%
  \BibitemOpen
  \bibfield  {author} {\bibinfo {author} {\bibfnamefont {X.}~\bibnamefont
  {Deng}}, \bibinfo {author} {\bibfnamefont {S.}~\bibnamefont {Ray}}, \bibinfo
  {author} {\bibfnamefont {S.}~\bibnamefont {Sinha}}, \bibinfo {author}
  {\bibfnamefont {G.~V.}\ \bibnamefont {Shlyapnikov}}, \ and\ \bibinfo {author}
  {\bibfnamefont {L.}~\bibnamefont {Santos}},\ }\href {\doibase
  10.1103/PhysRevLett.123.025301} {\bibfield  {journal} {\bibinfo  {journal}
  {Phys. Rev. Lett.}\ }\textbf {\bibinfo {volume} {123}},\ \bibinfo {pages}
  {025301} (\bibinfo {year} {2019})}\BibitemShut {NoStop}%
\bibitem [{\citenamefont {Yao}\ \emph {et~al.}(2019)\citenamefont {Yao},
  \citenamefont {Khoudli}, \citenamefont {Bresque},\ and\ \citenamefont
  {Sanchez-Palencia}}]{Hepeng19}%
  \BibitemOpen
  \bibfield  {author} {\bibinfo {author} {\bibfnamefont {H.}~\bibnamefont
  {Yao}}, \bibinfo {author} {\bibfnamefont {H.}~\bibnamefont {Khoudli}},
  \bibinfo {author} {\bibfnamefont {L.}~\bibnamefont {Bresque}}, \ and\
  \bibinfo {author} {\bibfnamefont {L.}~\bibnamefont {Sanchez-Palencia}},\
  }\href {\doibase 10.1103/PhysRevLett.123.070405} {\bibfield  {journal}
  {\bibinfo  {journal} {Phys. Rev. Lett.}\ }\textbf {\bibinfo {volume} {123}},\
  \bibinfo {pages} {070405} (\bibinfo {year} {2019})}\BibitemShut {NoStop}%
\bibitem [{\citenamefont {Liu}\ \emph {et~al.}(2022)\citenamefont {Liu},
  \citenamefont {Xia}, \citenamefont {Longhi},\ and\ \citenamefont
  {Sanchez-Palencia}}]{Liu_22}%
  \BibitemOpen
  \bibfield  {author} {\bibinfo {author} {\bibfnamefont {T.}~\bibnamefont
  {Liu}}, \bibinfo {author} {\bibfnamefont {X.}~\bibnamefont {Xia}}, \bibinfo
  {author} {\bibfnamefont {S.}~\bibnamefont {Longhi}}, \ and\ \bibinfo {author}
  {\bibfnamefont {L.}~\bibnamefont {Sanchez-Palencia}},\ }\href {\doibase
  10.21468/SciPostPhys.12.1.027} {\bibfield  {journal} {\bibinfo  {journal}
  {SciPost Phys.}\ }\textbf {\bibinfo {volume} {12}},\ \bibinfo {pages} {27}
  (\bibinfo {year} {2022})}\BibitemShut {NoStop}%
\bibitem [{\citenamefont {Wang}\ \emph {et~al.}(2015)\citenamefont {Wang},
  \citenamefont {Lian},\ and\ \citenamefont {Zhang}}]{wang2015}%
  \BibitemOpen
  \bibfield  {author} {\bibinfo {author} {\bibfnamefont {J.}~\bibnamefont
  {Wang}}, \bibinfo {author} {\bibfnamefont {B.}~\bibnamefont {Lian}}, \ and\
  \bibinfo {author} {\bibfnamefont {S.-C.}\ \bibnamefont {Zhang}},\ }\href
  {\doibase 10.1103/PhysRevLett.115.036805} {\bibfield  {journal} {\bibinfo
  {journal} {Phys. Rev. Lett.}\ }\textbf {\bibinfo {volume} {115}},\ \bibinfo
  {pages} {036805} (\bibinfo {year} {2015})}\BibitemShut {NoStop}%
\bibitem [{Sup()}]{SuppMater}%
  \BibitemOpen
  \href@noop {} {\ }\bibinfo {note} {See Supplemental Material at XXXX-XXXX for
  the details on the QSHI models, reservoir effect, connection between phase
  diagrams and band structures, normalized participation ratio, Born
  approximation for the continuum and lattice models.}\BibitemShut {Stop}%
\bibitem [{\citenamefont {Landauer}(1970)}]{Landauer1970}%
  \BibitemOpen
  \bibfield  {author} {\bibinfo {author} {\bibfnamefont {R.}~\bibnamefont
  {Landauer}},\ }\href {\doibase 10.1080/14786437008238472} {\bibfield
  {journal} {\bibinfo  {journal} {Philos. Mag.}\ }\textbf {\bibinfo {volume}
  {21}},\ \bibinfo {pages} {863} (\bibinfo {year} {1970})}\BibitemShut
  {NoStop}%
\bibitem [{\citenamefont {B\"uttiker}(1988)}]{Buttiker1988}%
  \BibitemOpen
  \bibfield  {author} {\bibinfo {author} {\bibfnamefont {M.}~\bibnamefont
  {B\"uttiker}},\ }\href {\doibase 10.1103/PhysRevB.38.9375} {\bibfield
  {journal} {\bibinfo  {journal} {Phys. Rev. B}\ }\textbf {\bibinfo {volume}
  {38}},\ \bibinfo {pages} {9375} (\bibinfo {year} {1988})}\BibitemShut
  {NoStop}%
\bibitem [{\citenamefont {Rotter}\ \emph {et~al.}(2000)\citenamefont {Rotter},
  \citenamefont {Tang}, \citenamefont {Wirtz}, \citenamefont {Trost},\ and\
  \citenamefont {Burgd\"orfer}}]{Rotter00}%
  \BibitemOpen
  \bibfield  {author} {\bibinfo {author} {\bibfnamefont {S.}~\bibnamefont
  {Rotter}}, \bibinfo {author} {\bibfnamefont {J.-Z.}\ \bibnamefont {Tang}},
  \bibinfo {author} {\bibfnamefont {L.}~\bibnamefont {Wirtz}}, \bibinfo
  {author} {\bibfnamefont {J.}~\bibnamefont {Trost}}, \ and\ \bibinfo {author}
  {\bibfnamefont {J.}~\bibnamefont {Burgd\"orfer}},\ }\href {\doibase
  10.1103/PhysRevB.62.1950} {\bibfield  {journal} {\bibinfo  {journal} {Phys.
  Rev. B}\ }\textbf {\bibinfo {volume} {62}},\ \bibinfo {pages} {1950}
  (\bibinfo {year} {2000})}\BibitemShut {NoStop}%
\bibitem [{\citenamefont {Rotter}\ \emph {et~al.}(2003)\citenamefont {Rotter},
  \citenamefont {Weingartner}, \citenamefont {Rohringer},\ and\ \citenamefont
  {Burgd\"orfer}}]{Rotter03}%
  \BibitemOpen
  \bibfield  {author} {\bibinfo {author} {\bibfnamefont {S.}~\bibnamefont
  {Rotter}}, \bibinfo {author} {\bibfnamefont {B.}~\bibnamefont {Weingartner}},
  \bibinfo {author} {\bibfnamefont {N.}~\bibnamefont {Rohringer}}, \ and\
  \bibinfo {author} {\bibfnamefont {J.}~\bibnamefont {Burgd\"orfer}},\ }\href
  {\doibase 10.1103/PhysRevB.68.165302} {\bibfield  {journal} {\bibinfo
  {journal} {Phys. Rev. B}\ }\textbf {\bibinfo {volume} {68}},\ \bibinfo
  {pages} {165302} (\bibinfo {year} {2003})}\BibitemShut {NoStop}%
\bibitem [{\citenamefont {Libisch}\ \emph {et~al.}(2012)\citenamefont
  {Libisch}, \citenamefont {Rotter},\ and\ \citenamefont
  {Burgdörfer}}]{Libisch_2012}%
  \BibitemOpen
  \bibfield  {author} {\bibinfo {author} {\bibfnamefont {F.}~\bibnamefont
  {Libisch}}, \bibinfo {author} {\bibfnamefont {S.}~\bibnamefont {Rotter}}, \
  and\ \bibinfo {author} {\bibfnamefont {J.}~\bibnamefont {Burgdörfer}},\
  }\href {\doibase 10.1088/1367-2630/14/12/123006} {\bibfield  {journal}
  {\bibinfo  {journal} {New Journal of Physics}\ }\textbf {\bibinfo {volume}
  {14}},\ \bibinfo {pages} {123006} (\bibinfo {year} {2012})}\BibitemShut
  {NoStop}%
\bibitem [{\citenamefont {Zimmermann}\ and\ \citenamefont
  {Schindler}(2009)}]{Zimmermann09}%
  \BibitemOpen
  \bibfield  {author} {\bibinfo {author} {\bibfnamefont {R.}~\bibnamefont
  {Zimmermann}}\ and\ \bibinfo {author} {\bibfnamefont {C.}~\bibnamefont
  {Schindler}},\ }\href {\doibase 10.1103/PhysRevB.80.144202} {\bibfield
  {journal} {\bibinfo  {journal} {Phys. Rev. B}\ }\textbf {\bibinfo {volume}
  {80}},\ \bibinfo {pages} {144202} (\bibinfo {year} {2009})}\BibitemShut
  {NoStop}%
\bibitem [{\citenamefont {{Chu, Rui-Lin}}\ \emph {et~al.}(2012)\citenamefont
  {{Chu, Rui-Lin}}, \citenamefont {{Lu, Jie}},\ and\ \citenamefont {{Shen,
  Shun-Qing}}}]{refId0}%
  \BibitemOpen
  \bibfield  {author} {\bibinfo {author} {\bibnamefont {{Chu, Rui-Lin}}},
  \bibinfo {author} {\bibnamefont {{Lu, Jie}}}, \ and\ \bibinfo {author}
  {\bibnamefont {{Shen, Shun-Qing}}},\ }\href {\doibase
  10.1209/0295-5075/100/17013} {\bibfield  {journal} {\bibinfo  {journal}
  {EPL}\ }\textbf {\bibinfo {volume} {100}},\ \bibinfo {pages} {17013}
  (\bibinfo {year} {2012})}\BibitemShut {NoStop}%
\bibitem [{\citenamefont {Hashimoto}\ \emph {et~al.}(2008)\citenamefont
  {Hashimoto}, \citenamefont {Sohrmann}, \citenamefont {Wiebe}, \citenamefont
  {Inaoka}, \citenamefont {Meier}, \citenamefont {Hirayama}, \citenamefont
  {R\"omer}, \citenamefont {Wiesendanger},\ and\ \citenamefont
  {Morgenstern}}]{Hashimoto08}%
  \BibitemOpen
  \bibfield  {author} {\bibinfo {author} {\bibfnamefont {K.}~\bibnamefont
  {Hashimoto}}, \bibinfo {author} {\bibfnamefont {C.}~\bibnamefont {Sohrmann}},
  \bibinfo {author} {\bibfnamefont {J.}~\bibnamefont {Wiebe}}, \bibinfo
  {author} {\bibfnamefont {T.}~\bibnamefont {Inaoka}}, \bibinfo {author}
  {\bibfnamefont {F.}~\bibnamefont {Meier}}, \bibinfo {author} {\bibfnamefont
  {Y.}~\bibnamefont {Hirayama}}, \bibinfo {author} {\bibfnamefont {R.~A.}\
  \bibnamefont {R\"omer}}, \bibinfo {author} {\bibfnamefont {R.}~\bibnamefont
  {Wiesendanger}}, \ and\ \bibinfo {author} {\bibfnamefont {M.}~\bibnamefont
  {Morgenstern}},\ }\href {\doibase 10.1103/PhysRevLett.101.256802} {\bibfield
  {journal} {\bibinfo  {journal} {Phys. Rev. Lett.}\ }\textbf {\bibinfo
  {volume} {101}},\ \bibinfo {pages} {256802} (\bibinfo {year}
  {2008})}\BibitemShut {NoStop}%
\bibitem [{\citenamefont {Huckestein}(1995)}]{Huckestein95}%
  \BibitemOpen
  \bibfield  {author} {\bibinfo {author} {\bibfnamefont {B.}~\bibnamefont
  {Huckestein}},\ }\href {\doibase 10.1103/RevModPhys.67.357} {\bibfield
  {journal} {\bibinfo  {journal} {Rev. Mod. Phys.}\ }\textbf {\bibinfo {volume}
  {67}},\ \bibinfo {pages} {357} (\bibinfo {year} {1995})}\BibitemShut
  {NoStop}%
\bibitem [{\citenamefont {Tokura}\ \emph {et~al.}(2019)\citenamefont {Tokura},
  \citenamefont {Yasuda},\ and\ \citenamefont
  {Tsukazaki}}]{tokura2019magnetic}%
  \BibitemOpen
  \bibfield  {author} {\bibinfo {author} {\bibfnamefont {Y.}~\bibnamefont
  {Tokura}}, \bibinfo {author} {\bibfnamefont {K.}~\bibnamefont {Yasuda}}, \
  and\ \bibinfo {author} {\bibfnamefont {A.}~\bibnamefont {Tsukazaki}},\
  }\href@noop {} {\bibfield  {journal} {\bibinfo  {journal} {Nature Reviews
  Physics}\ }\textbf {\bibinfo {volume} {1}},\ \bibinfo {pages} {126} (\bibinfo
  {year} {2019})}\BibitemShut {NoStop}%
\bibitem [{\citenamefont {Watanabe}\ \emph {et~al.}(2019)\citenamefont
  {Watanabe}, \citenamefont {Yoshimi}, \citenamefont {Kawamura}, \citenamefont
  {Mogi}, \citenamefont {Tsukazaki}, \citenamefont {Yu}, \citenamefont
  {Nakajima}, \citenamefont {Takahashi}, \citenamefont {Kawasaki},\ and\
  \citenamefont {Tokura}}]{watanabe2019quantum}%
  \BibitemOpen
  \bibfield  {author} {\bibinfo {author} {\bibfnamefont {R.}~\bibnamefont
  {Watanabe}}, \bibinfo {author} {\bibfnamefont {R.}~\bibnamefont {Yoshimi}},
  \bibinfo {author} {\bibfnamefont {M.}~\bibnamefont {Kawamura}}, \bibinfo
  {author} {\bibfnamefont {M.}~\bibnamefont {Mogi}}, \bibinfo {author}
  {\bibfnamefont {A.}~\bibnamefont {Tsukazaki}}, \bibinfo {author}
  {\bibfnamefont {X.}~\bibnamefont {Yu}}, \bibinfo {author} {\bibfnamefont
  {K.}~\bibnamefont {Nakajima}}, \bibinfo {author} {\bibfnamefont {K.~S.}\
  \bibnamefont {Takahashi}}, \bibinfo {author} {\bibfnamefont {M.}~\bibnamefont
  {Kawasaki}}, \ and\ \bibinfo {author} {\bibfnamefont {Y.}~\bibnamefont
  {Tokura}},\ }\href@noop {} {\bibfield  {journal} {\bibinfo  {journal}
  {Applied Physics Letters}\ }\textbf {\bibinfo {volume} {115}},\ \bibinfo
  {pages} {102403} (\bibinfo {year} {2019})}\BibitemShut {NoStop}%
\bibitem [{\citenamefont {Satake}\ \emph {et~al.}(2020)\citenamefont {Satake},
  \citenamefont {Shiogai}, \citenamefont {Mazur}, \citenamefont {Kimura},
  \citenamefont {Awaji}, \citenamefont {Fujiwara}, \citenamefont {Nojima},
  \citenamefont {Nomura}, \citenamefont {Souma}, \citenamefont {Sato},
  \citenamefont {Dietl},\ and\ \citenamefont {Tsukazaki}}]{Satake20}%
  \BibitemOpen
  \bibfield  {author} {\bibinfo {author} {\bibfnamefont {Y.}~\bibnamefont
  {Satake}}, \bibinfo {author} {\bibfnamefont {J.}~\bibnamefont {Shiogai}},
  \bibinfo {author} {\bibfnamefont {G.~P.}\ \bibnamefont {Mazur}}, \bibinfo
  {author} {\bibfnamefont {S.}~\bibnamefont {Kimura}}, \bibinfo {author}
  {\bibfnamefont {S.}~\bibnamefont {Awaji}}, \bibinfo {author} {\bibfnamefont
  {K.}~\bibnamefont {Fujiwara}}, \bibinfo {author} {\bibfnamefont
  {T.}~\bibnamefont {Nojima}}, \bibinfo {author} {\bibfnamefont
  {K.}~\bibnamefont {Nomura}}, \bibinfo {author} {\bibfnamefont
  {S.}~\bibnamefont {Souma}}, \bibinfo {author} {\bibfnamefont
  {T.}~\bibnamefont {Sato}}, \bibinfo {author} {\bibfnamefont {T.}~\bibnamefont
  {Dietl}}, \ and\ \bibinfo {author} {\bibfnamefont {A.}~\bibnamefont
  {Tsukazaki}},\ }\href {\doibase 10.1103/PhysRevMaterials.4.044202} {\bibfield
   {journal} {\bibinfo  {journal} {Phys. Rev. Materials}\ }\textbf {\bibinfo
  {volume} {4}},\ \bibinfo {pages} {044202} (\bibinfo {year}
  {2020})}\BibitemShut {NoStop}%
\bibitem [{\citenamefont {B\'eri}\ and\ \citenamefont
  {Cooper}(2011)}]{cooper11}%
  \BibitemOpen
  \bibfield  {author} {\bibinfo {author} {\bibfnamefont {B.}~\bibnamefont
  {B\'eri}}\ and\ \bibinfo {author} {\bibfnamefont {N.~R.}\ \bibnamefont
  {Cooper}},\ }\href {\doibase 10.1103/PhysRevLett.107.145301} {\bibfield
  {journal} {\bibinfo  {journal} {Phys. Rev. Lett.}\ }\textbf {\bibinfo
  {volume} {107}},\ \bibinfo {pages} {145301} (\bibinfo {year}
  {2011})}\BibitemShut {NoStop}%
\bibitem [{\citenamefont {Aidelsburger}\ \emph {et~al.}(2013)\citenamefont
  {Aidelsburger}, \citenamefont {Atala}, \citenamefont {Lohse}, \citenamefont
  {Barreiro}, \citenamefont {Paredes},\ and\ \citenamefont
  {Bloch}}]{Aidelsburger13}%
  \BibitemOpen
  \bibfield  {author} {\bibinfo {author} {\bibfnamefont {M.}~\bibnamefont
  {Aidelsburger}}, \bibinfo {author} {\bibfnamefont {M.}~\bibnamefont {Atala}},
  \bibinfo {author} {\bibfnamefont {M.}~\bibnamefont {Lohse}}, \bibinfo
  {author} {\bibfnamefont {J.~T.}\ \bibnamefont {Barreiro}}, \bibinfo {author}
  {\bibfnamefont {B.}~\bibnamefont {Paredes}}, \ and\ \bibinfo {author}
  {\bibfnamefont {I.}~\bibnamefont {Bloch}},\ }\href {\doibase
  10.1103/PhysRevLett.111.185301} {\bibfield  {journal} {\bibinfo  {journal}
  {Phys. Rev. Lett.}\ }\textbf {\bibinfo {volume} {111}},\ \bibinfo {pages}
  {185301} (\bibinfo {year} {2013})}\BibitemShut {NoStop}%
\bibitem [{\citenamefont {Atala}\ \emph {et~al.}(2014)\citenamefont {Atala},
  \citenamefont {Aidelsburger}, \citenamefont {Lohse}, \citenamefont
  {Barreiro}, \citenamefont {Paredes},\ and\ \citenamefont
  {Bloch}}]{atala2014observation}%
  \BibitemOpen
  \bibfield  {author} {\bibinfo {author} {\bibfnamefont {M.}~\bibnamefont
  {Atala}}, \bibinfo {author} {\bibfnamefont {M.}~\bibnamefont {Aidelsburger}},
  \bibinfo {author} {\bibfnamefont {M.}~\bibnamefont {Lohse}}, \bibinfo
  {author} {\bibfnamefont {J.~T.}\ \bibnamefont {Barreiro}}, \bibinfo {author}
  {\bibfnamefont {B.}~\bibnamefont {Paredes}}, \ and\ \bibinfo {author}
  {\bibfnamefont {I.}~\bibnamefont {Bloch}},\ }\href@noop {} {\bibfield
  {journal} {\bibinfo  {journal} {Nature Physics}\ }\textbf {\bibinfo {volume}
  {10}},\ \bibinfo {pages} {588} (\bibinfo {year} {2014})}\BibitemShut
  {NoStop}%
\bibitem [{\citenamefont {Aidelsburger}\ \emph {et~al.}(2015)\citenamefont
  {Aidelsburger}, \citenamefont {Lohse}, \citenamefont {Schweizer},
  \citenamefont {Atala}, \citenamefont {Barreiro}, \citenamefont
  {Nascimb{\`e}ne}, \citenamefont {Cooper}, \citenamefont {Bloch},\ and\
  \citenamefont {Goldman}}]{aidelsburger2015measuring}%
  \BibitemOpen
  \bibfield  {author} {\bibinfo {author} {\bibfnamefont {M.}~\bibnamefont
  {Aidelsburger}}, \bibinfo {author} {\bibfnamefont {M.}~\bibnamefont {Lohse}},
  \bibinfo {author} {\bibfnamefont {C.}~\bibnamefont {Schweizer}}, \bibinfo
  {author} {\bibfnamefont {M.}~\bibnamefont {Atala}}, \bibinfo {author}
  {\bibfnamefont {J.~T.}\ \bibnamefont {Barreiro}}, \bibinfo {author}
  {\bibfnamefont {S.}~\bibnamefont {Nascimb{\`e}ne}}, \bibinfo {author}
  {\bibfnamefont {N.}~\bibnamefont {Cooper}}, \bibinfo {author} {\bibfnamefont
  {I.}~\bibnamefont {Bloch}}, \ and\ \bibinfo {author} {\bibfnamefont
  {N.}~\bibnamefont {Goldman}},\ }\href@noop {} {\bibfield  {journal} {\bibinfo
   {journal} {Nature Physics}\ }\textbf {\bibinfo {volume} {11}},\ \bibinfo
  {pages} {162} (\bibinfo {year} {2015})}\BibitemShut {NoStop}%
\bibitem [{\citenamefont {Jotzu}\ \emph {et~al.}(2014)\citenamefont {Jotzu},
  \citenamefont {Messer}, \citenamefont {Desbuquois}, \citenamefont {Lebrat},
  \citenamefont {Uehlinger}, \citenamefont {Greif},\ and\ \citenamefont
  {Esslinger}}]{jotzu2014experimental}%
  \BibitemOpen
  \bibfield  {author} {\bibinfo {author} {\bibfnamefont {G.}~\bibnamefont
  {Jotzu}}, \bibinfo {author} {\bibfnamefont {M.}~\bibnamefont {Messer}},
  \bibinfo {author} {\bibfnamefont {R.}~\bibnamefont {Desbuquois}}, \bibinfo
  {author} {\bibfnamefont {M.}~\bibnamefont {Lebrat}}, \bibinfo {author}
  {\bibfnamefont {T.}~\bibnamefont {Uehlinger}}, \bibinfo {author}
  {\bibfnamefont {D.}~\bibnamefont {Greif}}, \ and\ \bibinfo {author}
  {\bibfnamefont {T.}~\bibnamefont {Esslinger}},\ }\href@noop {} {\bibfield
  {journal} {\bibinfo  {journal} {Nature}\ }\textbf {\bibinfo {volume} {515}},\
  \bibinfo {pages} {237} (\bibinfo {year} {2014})}\BibitemShut {NoStop}%
\bibitem [{\citenamefont {Fl{\"a}schner}\ \emph {et~al.}(2016)\citenamefont
  {Fl{\"a}schner}, \citenamefont {Rem}, \citenamefont {Tarnowski},
  \citenamefont {Vogel}, \citenamefont {L{\"u}hmann}, \citenamefont
  {Sengstock},\ and\ \citenamefont {Weitenberg}}]{flaschner2016experimental}%
  \BibitemOpen
  \bibfield  {author} {\bibinfo {author} {\bibfnamefont {N.}~\bibnamefont
  {Fl{\"a}schner}}, \bibinfo {author} {\bibfnamefont {B.}~\bibnamefont {Rem}},
  \bibinfo {author} {\bibfnamefont {M.}~\bibnamefont {Tarnowski}}, \bibinfo
  {author} {\bibfnamefont {D.}~\bibnamefont {Vogel}}, \bibinfo {author}
  {\bibfnamefont {D.-S.}\ \bibnamefont {L{\"u}hmann}}, \bibinfo {author}
  {\bibfnamefont {K.}~\bibnamefont {Sengstock}}, \ and\ \bibinfo {author}
  {\bibfnamefont {C.}~\bibnamefont {Weitenberg}},\ }\href@noop {} {\bibfield
  {journal} {\bibinfo  {journal} {Science}\ }\textbf {\bibinfo {volume}
  {352}},\ \bibinfo {pages} {1091} (\bibinfo {year} {2016})}\BibitemShut
  {NoStop}%
\bibitem [{\citenamefont {Su}\ \emph {et~al.}(1979)\citenamefont {Su},
  \citenamefont {Schrieffer},\ and\ \citenamefont {Heeger}}]{SSH_79}%
  \BibitemOpen
  \bibfield  {author} {\bibinfo {author} {\bibfnamefont {W.~P.}\ \bibnamefont
  {Su}}, \bibinfo {author} {\bibfnamefont {J.~R.}\ \bibnamefont {Schrieffer}},
  \ and\ \bibinfo {author} {\bibfnamefont {A.~J.}\ \bibnamefont {Heeger}},\
  }\href {\doibase 10.1103/PhysRevLett.42.1698} {\bibfield  {journal} {\bibinfo
   {journal} {Phys. Rev. Lett.}\ }\textbf {\bibinfo {volume} {42}},\ \bibinfo
  {pages} {1698} (\bibinfo {year} {1979})}\BibitemShut {NoStop}%
\bibitem [{\citenamefont {Atala}\ \emph {et~al.}(2013)\citenamefont {Atala},
  \citenamefont {Aidelsburger}, \citenamefont {Barreiro}, \citenamefont
  {Abanin}, \citenamefont {Kitagawa}, \citenamefont {Demler},\ and\
  \citenamefont {Bloch}}]{atala2013direct}%
  \BibitemOpen
  \bibfield  {author} {\bibinfo {author} {\bibfnamefont {M.}~\bibnamefont
  {Atala}}, \bibinfo {author} {\bibfnamefont {M.}~\bibnamefont {Aidelsburger}},
  \bibinfo {author} {\bibfnamefont {J.~T.}\ \bibnamefont {Barreiro}}, \bibinfo
  {author} {\bibfnamefont {D.}~\bibnamefont {Abanin}}, \bibinfo {author}
  {\bibfnamefont {T.}~\bibnamefont {Kitagawa}}, \bibinfo {author}
  {\bibfnamefont {E.}~\bibnamefont {Demler}}, \ and\ \bibinfo {author}
  {\bibfnamefont {I.}~\bibnamefont {Bloch}},\ }\href@noop {} {\bibfield
  {journal} {\bibinfo  {journal} {Nature Physics}\ }\textbf {\bibinfo {volume}
  {9}},\ \bibinfo {pages} {795} (\bibinfo {year} {2013})}\BibitemShut {NoStop}%
\bibitem [{\citenamefont {Guidoni}\ \emph {et~al.}(1997)\citenamefont
  {Guidoni}, \citenamefont {Trich\'e}, \citenamefont {Verkerk},\ and\
  \citenamefont {Grynberg}}]{Guidoni97}%
  \BibitemOpen
  \bibfield  {author} {\bibinfo {author} {\bibfnamefont {L.}~\bibnamefont
  {Guidoni}}, \bibinfo {author} {\bibfnamefont {C.}~\bibnamefont {Trich\'e}},
  \bibinfo {author} {\bibfnamefont {P.}~\bibnamefont {Verkerk}}, \ and\
  \bibinfo {author} {\bibfnamefont {G.}~\bibnamefont {Grynberg}},\ }\href
  {\doibase 10.1103/PhysRevLett.79.3363} {\bibfield  {journal} {\bibinfo
  {journal} {Phys. Rev. Lett.}\ }\textbf {\bibinfo {volume} {79}},\ \bibinfo
  {pages} {3363} (\bibinfo {year} {1997})}\BibitemShut {NoStop}%
\bibitem [{\citenamefont {Roati}\ \emph {et~al.}(2008)\citenamefont {Roati},
  \citenamefont {D’Errico}, \citenamefont {Fallani}, \citenamefont {Fattori},
  \citenamefont {Fort}, \citenamefont {Zaccanti}, \citenamefont {Modugno},
  \citenamefont {Modugno},\ and\ \citenamefont {Inguscio}}]{roati2008anderson}%
  \BibitemOpen
  \bibfield  {author} {\bibinfo {author} {\bibfnamefont {G.}~\bibnamefont
  {Roati}}, \bibinfo {author} {\bibfnamefont {C.}~\bibnamefont {D’Errico}},
  \bibinfo {author} {\bibfnamefont {L.}~\bibnamefont {Fallani}}, \bibinfo
  {author} {\bibfnamefont {M.}~\bibnamefont {Fattori}}, \bibinfo {author}
  {\bibfnamefont {C.}~\bibnamefont {Fort}}, \bibinfo {author} {\bibfnamefont
  {M.}~\bibnamefont {Zaccanti}}, \bibinfo {author} {\bibfnamefont
  {G.}~\bibnamefont {Modugno}}, \bibinfo {author} {\bibfnamefont
  {M.}~\bibnamefont {Modugno}}, \ and\ \bibinfo {author} {\bibfnamefont
  {M.}~\bibnamefont {Inguscio}},\ }\href@noop {} {\bibfield  {journal}
  {\bibinfo  {journal} {Nature}\ }\textbf {\bibinfo {volume} {453}},\ \bibinfo
  {pages} {895} (\bibinfo {year} {2008})}\BibitemShut {NoStop}%
\bibitem [{\citenamefont {Billy}\ \emph {et~al.}(2008)\citenamefont {Billy},
  \citenamefont {Josse}, \citenamefont {Zuo}, \citenamefont {Bernard},
  \citenamefont {Hambrecht}, \citenamefont {Lugan}, \citenamefont
  {Cl{\'e}ment}, \citenamefont {Sanchez-Palencia}, \citenamefont {Bouyer},\
  and\ \citenamefont {Aspect}}]{billy2008direct}%
  \BibitemOpen
  \bibfield  {author} {\bibinfo {author} {\bibfnamefont {J.}~\bibnamefont
  {Billy}}, \bibinfo {author} {\bibfnamefont {V.}~\bibnamefont {Josse}},
  \bibinfo {author} {\bibfnamefont {Z.}~\bibnamefont {Zuo}}, \bibinfo {author}
  {\bibfnamefont {A.}~\bibnamefont {Bernard}}, \bibinfo {author} {\bibfnamefont
  {B.}~\bibnamefont {Hambrecht}}, \bibinfo {author} {\bibfnamefont
  {P.}~\bibnamefont {Lugan}}, \bibinfo {author} {\bibfnamefont
  {D.}~\bibnamefont {Cl{\'e}ment}}, \bibinfo {author} {\bibfnamefont
  {L.}~\bibnamefont {Sanchez-Palencia}}, \bibinfo {author} {\bibfnamefont
  {P.}~\bibnamefont {Bouyer}}, \ and\ \bibinfo {author} {\bibfnamefont
  {A.}~\bibnamefont {Aspect}},\ }\href@noop {} {\bibfield  {journal} {\bibinfo
  {journal} {Nature}\ }\textbf {\bibinfo {volume} {453}},\ \bibinfo {pages}
  {891} (\bibinfo {year} {2008})}\BibitemShut {NoStop}%
\bibitem [{\citenamefont {Lye}\ \emph {et~al.}(2005)\citenamefont {Lye},
  \citenamefont {Fallani}, \citenamefont {Modugno}, \citenamefont {Wiersma},
  \citenamefont {Fort},\ and\ \citenamefont {Inguscio}}]{Lye05}%
  \BibitemOpen
  \bibfield  {author} {\bibinfo {author} {\bibfnamefont {J.~E.}\ \bibnamefont
  {Lye}}, \bibinfo {author} {\bibfnamefont {L.}~\bibnamefont {Fallani}},
  \bibinfo {author} {\bibfnamefont {M.}~\bibnamefont {Modugno}}, \bibinfo
  {author} {\bibfnamefont {D.~S.}\ \bibnamefont {Wiersma}}, \bibinfo {author}
  {\bibfnamefont {C.}~\bibnamefont {Fort}}, \ and\ \bibinfo {author}
  {\bibfnamefont {M.}~\bibnamefont {Inguscio}},\ }\href {\doibase
  10.1103/PhysRevLett.95.070401} {\bibfield  {journal} {\bibinfo  {journal}
  {Phys. Rev. Lett.}\ }\textbf {\bibinfo {volume} {95}},\ \bibinfo {pages}
  {070401} (\bibinfo {year} {2005})}\BibitemShut {NoStop}%
\end{thebibliography}%

\clearpage

\begin{onecolumngrid}
	
\begin{center}
	{\fontsize{12}{12}\selectfont
		\textbf{Supplemental Materials: Correlated disorder induced anomalous transport in time reversal symmetry breaking topological insulator\\[5mm]}}
	{\normalsize Takuya Okugawa,$^{1}$ Tanay Nag,$^{1}$ and Dante M.\ Kennes$^{1,2}$\\[1mm]}
	{\small $^1$\textit{Institut f\"ur Theorie der Statistischen Physik, RWTH Aachen, 
52056 Aachen, Germany and JARA - Fundamentals of Future Information Technology}\\[0.5mm]}
	{\small $^2$\textit{Max Planck Institute for the Structure and Dynamics of Matter, Center for Free Electron Laser Science, 22761 Hamburg, Germany}\\[0.5mm]}
	{}
\end{center}

\title{Supplemental Materials: Correlated disorder induced anomalous transport in time reversal symmetry breaking topological insulator }  
\author{Takuya Okugawa}
\affiliation{Institut f\"ur Theorie der Statistischen Physik, RWTH Aachen, 
52056 Aachen, Germany and JARA - Fundamentals of Future Information Technology.}

\author{Tanay Nag}
\email{tnag@physik.rwth-aachen.de}
\affiliation{Institut f\"ur Theorie der Statistischen Physik, RWTH Aachen, 
52056 Aachen, Germany and JARA - Fundamentals of Future Information Technology.}

\author{Dante M.\ Kennes}
\email{dkennes@rwth-aachen.de}
\affiliation{Institut f\"ur Theorie der Statistischen Physik, RWTH Aachen, 
52056 Aachen, Germany and JARA - Fundamentals of Future Information Technology.}
\affiliation{Max Planck Institute for the Structure and Dynamics of Matter, Center for Free Electron Laser Science, 22761 Hamburg, Germany.}

\section{QSHI model}
In this section, we extensively discuss the QSHI model as given by  the Eq.~(1) of the main text. The model Hamiltonian for a QSHI is the following  \cite{Bernevig1757,wang2015} 
\begin{equation}
 H_0(\bm k)= {\bm N}\cdot {\bm \Gamma}=\sum^3_{i=1}N_i\Gamma_i
 \label{eq:model1}
\end{equation}
where $N_1=v_F \sin(k_y a)/a$, $N_2=-v_F \sin(k_x a)/a$, $N_3=m(\bm{k})=  m_0 + 2B[2- \cos(k_x a)-\cos(k_y a)]/a^2$ and $\Gamma_1= \tau_x \sigma_0$, $\Gamma_2= \tau_y \sigma_z $, and $\Gamma_3= \tau_z \sigma_0$. We note that ${\bm \tau} \in \{A,B \}$ and ${\bm \sigma}\in\{\uparrow,\downarrow \}$ represent orbital and spin degrees of freedom. 
Here, $v_F$ ($a$) denotes the Fermi velocity (lattice spacing).   
The QSHI model in Eq.~(\ref{eq:model1}) supports 
gapless helical edge modes, protected by TRS ${\mathcal T}=i \tau_z\sigma_x {\mathcal K}$ with ${\mathcal K}$ being the complex conjugate operator: ${\mathcal T}H_0(\bm k){\mathcal T}^{-1}=H_0(-\bm k)$. The model becomes trivially gapped for $m_0/B>0$ ($B$ is chosen to be positive). This model has unitary chiral symmetry and anti-unitary particle-hole symmetry, respectively, generated by   ${\mathcal C}=\tau_y \sigma_x$ and  ${\mathcal P}=\tau_x \sigma_z {\mathcal K}$: ${\mathcal C}H_0(\bm k){\mathcal C}^{-1}=-H_0(\bm k)$ and ${\mathcal P}H_0(\bm k){\mathcal P}^{-1}=-H_0(-\bm k)$.   Interestingly, the QSHI model  has mirror symmetry $M_{xy}={\mathcal C}_4 M_y$: $M_{xy} H_0(k_x,k_y,k_z)(M_{xy})^{-1}=H_0(k_y,k_x,k_z)$ where ${\mathcal C}_4$ [$M_y$] represents the generator of the four-fold rotational symmetry [mirror symmetry along $y$-axis].
A magnetic field, breaking TRS, can be introduced in the model, as discussed in Eq.~(2) of the main text: ${ H}(\bm k)=H_0(\bm k)+  gM \tau_z \sigma_z$  where $g$ being the Lande-g factor and $M$ is the magnetic exchange field. The chiral edge modes in ${ H}(\bm k)$ are preserved by the anti-unitary symmetry while the model does not have the unitary symmetry. 

\section{Reservoir effect}


\begin{figure*}[ht]
{\includegraphics[width=0.95\textwidth]{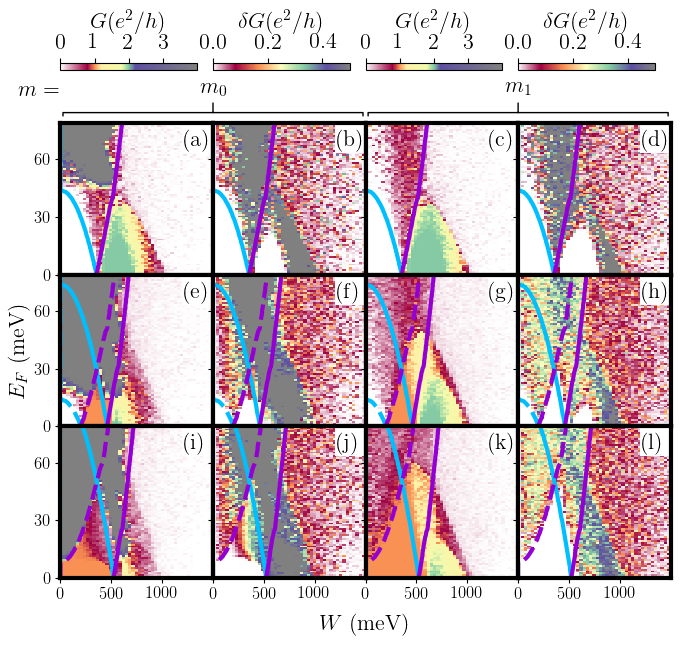}}
\caption{
(color online) (a), (e), and (i) [(c), (g), and (k)] depict the conductance $G$ with $gM=0$, $30$ and $52$ meV, respectively, for the longitudinal quasi-periodicity $\epsilon_i=W \cos(2 \pi \eta  i +\phi)/2$, considering the QSHI lead Hamiltonian $H_{L,R}(-|m_0|)$ ($m_0=44$ meV) [$H_{L,R}(-|m_1|)$ ($m_1=80 $ meV)]. The corresponding standard deviation $\delta G$ are shown for the longitudinal quasi-periodicity with $H_{L,R}(-|m_0|)$ [$H_{L,R}(-|m_1|)$] in (b), (f), and (j) [(d), (h), and (l)]. The system sizes are taken to be $L_x=400a$ for all panels and $L_y=100a$ for (a)-(d) and $L_y=200a$ for (e)-(l).
Notice that the central system is always described by the Hamiltonian $H_{\rm CS}(m_0,M,W)$ irrespective of the topological mass term in the leads. 
The QSCI phase does [does not] appears for $gM=52$ meV after QAHI phase for (k) [(i)] referring to the fact that the topological mass term in the QSHI leads is responsible for such phenomena. }
\label{fig:SM_fig1}
\end{figure*}

In this section, we analyze Fig. 3 (e) of the main text where the QSCI phase is no longer observed 
although the upper and lower block Hamiltonian $H_{u,l}(\bm k)$ both become topological with $\overline{m}_{0}^{u,l}<0$ as predicted by SCBA for disorder strength $W>550$ meV.   Such a counter-intuitive observations can be caused by a 
reservoir effect that we explain below at length.
In order to understand the underlying reason, we vary the topological mass term from $m_0$ to $m_1$ in the QSHI leads such that $|m_1|>|m_0|$ 
while keeping the Hamiltonian $H_{\rm CS}(m_0,M, W)$ of the central system unaltered (see Fig.~\ref{fig:SM_fig1}).

We concentrate on longitudinal quasi-periodicity $\epsilon_i=W \cos(2 \pi \eta i+\phi)/2$ with $\phi \in [0, 2\pi)$ in the central system. At first, we  consider a lead Hamiltonians given by $H_{L,R}(-|m_0|)$.
Following the SCBA, the QSCI phase is expected to show up after the QAHI phase when increasing $W$. In the conductance $G$, we observe such a behavior for $gM=30$ meV (see Fig.~\ref{fig:SM_fig1} (e)) while for strong magnetic field $gM=52$ meV (see Fig.~\ref{fig:SM_fig1} (i)), there is no signature of the QSCI phase following the QAHI phase upon increasing $W$.
We now change the lead Hamiltonians to 
$H_{L,R}(-|m_1|)$. 
In this case, we find the QAHI phase is followed by a QSCI phase for $gM=30$, and $52$ meV from the conductance analysis (see Figs.~\ref{fig:SM_fig1} (g) and (k)) which are in  accordance with the SCBA. 
Therefore, enhancing the topological mass ($|m_1|>|m_0|$) in the QSHI leads could resolve the apparent existence of a trivial  phase in conflict with $\overline{m}_{0}^{u,l}<0$ from SCBA.  Importantly, the robustness of the QSCI phase in Fig.~\ref{fig:SM_fig1} (g) is confirmed by $\delta G \to 0$ as shown in Fig.~\ref{fig:SM_fig1} (h). This is different from Fig.~\ref{fig:SM_fig1} (f) where $\delta G$ does not completely vanish inside the QSCI phase.

In addition, the QAHI phase with $gM=52$ meV is not accurately captured by the SCBA for $H_{L,R}(-|m_0|)$. We find a certain zone in the phase diagram with  $G<1$ inside the predicted QAHI phase as shown by the red patches. This zone is  bounded by the solid blue and purple dashed lines of SCBA, for $gM=52$ meV (see Fig.~\ref{fig:SM_fig1} (i)). These non-quantized patches vanish  when the QSHI leads are given by $H_{L,R}(-|m_1|)$ instead of $H_{L,R}(-|m_0|)$ complying with the SCBA for the above choice of the leads (see Fig.~\ref{fig:SM_fig1} (k)).

Note that when a topological mass term of higher magnitude is considered for the QSHI leads i.e., $|m_1|>|m_0|$,  the effects of the reservoir gets suppressed.  This effect may arise once the topological gap of the lead is less than or comparable to the gap of the central system.  
The interface between central system and leads essentially causes the bulk modes of the leads to hybridize with the edge modes of the central system resulting in  the contamination of the topological properties for the latter \cite{Okugawa20}. Hence, only by increasing the mass term in the QSHI leads, we recover the quantized edge transport of the central system more accurately. Upon inspecting the phase diagrams in Fig.~\ref{fig:SM_fig1}, one can comment that the topological gap in the leads $|m_1|$ has to be much larger than the renormalized bulk gap $\Delta$ of the central system ($|m_1|\gg|\Delta|$) to prevent bulk states of the leads from hybridizing with the edge modes in the central system.



\section{Phase diagrams and band structures}
\begin{figure}[ht]
	\centering
	\subfigure{\includegraphics[width=0.48\textwidth]{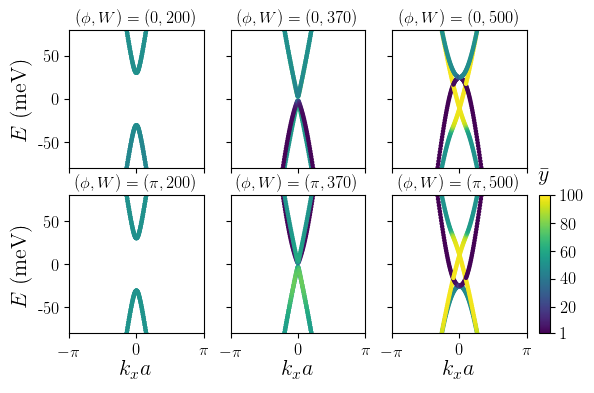}}
	\caption{(Color online) The energy dispersion of the non-magnetic isolated central system (Eq. (2) of the main text) in presence of transverse quasi-periodic potential ($\alpha=0$ and $\beta=1$) under the stripe geometry  with infinite (open) boundary condition along $x$ ($y$)-direction. Here, we depict the TPT between  the NI $\to$ QSHI phase i.e., $G=0 \to G=2$, as shown in Fig. 4 (a) of the main text.  We can find the $G=0$ phase for $W=200$, the TPT at $W=370$ and the $G=2$ phase for $W=500$. Moreover, we separately consider $\phi=0$ and $\pi$ to emphasize the emergent symmetry $E(k_x,W)=-E(-k_x,-W)$. The colorbar  denotes the average localization $\overline{y}$ of a given momentum mode at $k_x$ in the finite $y$-direction. 
	}
	\label{fig:SM_fig2}
\end{figure}

\begin{figure*}[ht]
	\centering
	\subfigure{\includegraphics[width=0.95\textwidth]{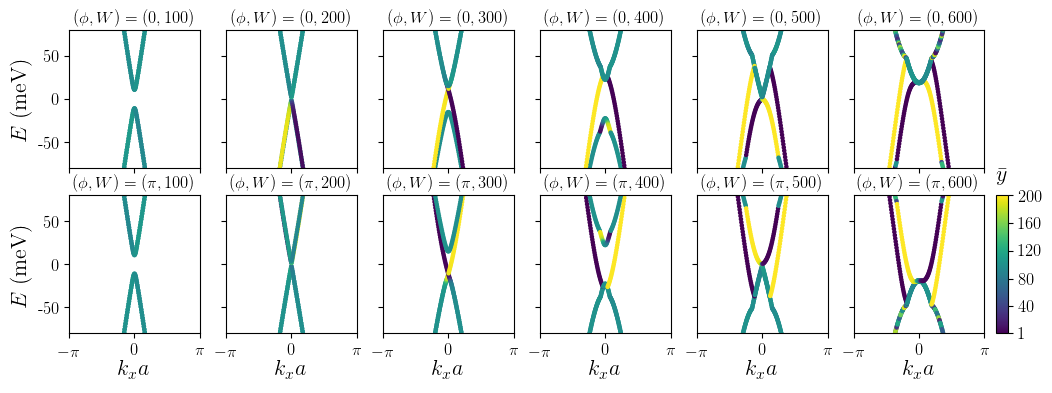}}
	\caption{(Color online) The energy dispersion of the isolated central system (Eq. (2) of the main text) for $gM=30$ meV in presence of transverse quasi-periodic potential ($\alpha=0$ and $\beta=1$) under the stripe geometry with infinite (open) boundary condition along $x$ ($y$)-direction. Here, we depict TPTs between the NI $\to$ QAHI phase i.e., $G=0 \to G=1$, 
	the QAHI $\to$ QSCI phase i.e., $G=1 \to G=2$,
	as shown in Fig. 4 (c) of the main text. We can find a $G=0$ phase for $W=100$, the TPT between NI and QAHI at $W=200$, $G=1$ phase for $W=300$ and $W=400$, the TPT between QAHI and QSCI phase at $W=500$, $G=2$ QSCI phase for $W=600$. Moreover, we separately consider $\phi=0$ and $\pi$ to emphasize the emergent symmetry $E(k_x,W)=-E(-k_x,-W)$. The color bar denotes the average localization of a given momentum mode at $k_x$ in the finite $y$-direction. 
	}
	\label{fig:SM_fig3}
\end{figure*}

\begin{figure}[ht]
	\centering
	\subfigure{\includegraphics[width=0.48\textwidth]{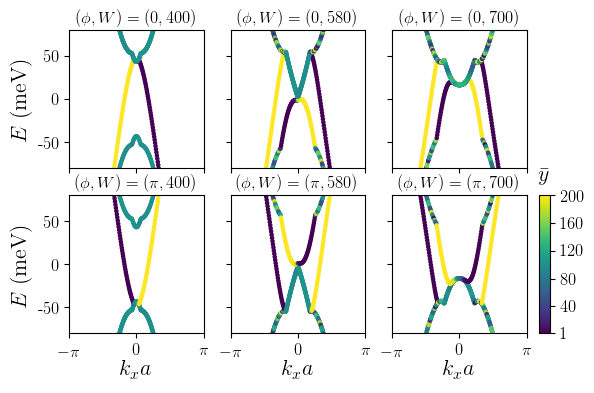}}
	\caption{(Color online) The energy dispersion of the isolated central system (Eq. (2) of the main text) for $gM=52$ meV in presence of transverse quasi-periodic potential ($\alpha=0$ and $\beta=1$) under the stripe geometry with infinite (open) boundary condition along $x$ ($y$)-direction. Here,we depict TPT between  the QAHI $\to$ QSCI phase i.e., $G=1 \to G=2$, as shown in Fig. 4 (e) of the main text.  We can find a $G=1$ phase for $W=400$, the TPT between $G=1$ and $G=2$ at $W=580$ and $G=2$ phase for $W=700$. Moreover, we separately consider $\phi=0$ and $\pi$ to emphasize the emergent symmetry $E(k_x,W)=-E(-k_x,-W)$.  
	}
	\label{fig:SM_fig4}
\end{figure}
Here, we discuss how one can understand the phase diagram for transverse quasi-periodicity which is shown in Fig. 4 of the main text, from the band structure of the isolated central system considering $L_x\to\infty$. In this case $k_x$ can be considered as a good quantum number due to the translation symmetry along the $x$-direction. Notice that the  quasi-periodic potential along the $y$-direction breaks the translation symmetry only along that direction. This enables us to probe the band structures by varying the disorder strength i.e., the amplitude of quasi-periodic potential $W$.  The results are shown in Figs.~\ref{fig:SM_fig2}, \ref{fig:SM_fig3} and \ref{fig:SM_fig4}, respectively, for $gM=0$, $30$ and $52$ meV.

We numerically diagonalize $H_{\rm CS}(m_0,M,W)= \sum_{j, j',k_x}[ {\mathcal H}_{j,j',k_x}(m_0,M) + \epsilon_j \delta_{j, j'}]C^{\dagger}_{j,k_x} C_{j',k_x} $ considering $\epsilon_j=W \cos(2 \pi \eta j + \phi)/2$ and $j=1, \cdots, L_y$ with $C= \{C_{A\uparrow},C_{A\downarrow},C_{B\uparrow},C_{B\downarrow}\}$. Here,  ${\mathcal H}_{j,j',k_x}(m_0,M)$ is obtained after the partial inverse Fourier transformation of ${ H}(\bm k)$ in Eq. (2) of the main text along the $y$-direction only. 
We further simplify the situation by considering only two specific values of $\phi=0$ and $\pi$ such that $\epsilon_j\to -\epsilon_j$ or equivalently $W \to -W$ for $\phi\to \phi+\pi$. This allows us to look for the correlation between the energy dispersion
and sign reversal in $W$ in a concrete manner. 
In addition, we measure average localization  of each momentum mode in the $y$-direction, associated with eigenenergy $E_n(k_x)$, as follows  $\bar y_n(k_x)=\sum_{j=1}^{L_y}  j(\sum_{q}|\psi_{n,q}(j,k_x)|^2)$  where $\psi_{n,q}(j,k_x)$ is the 
$j$-th component  of
$n$-th eigenstate in the basis $q= \{A\uparrow,A\downarrow,B\uparrow,B\downarrow\}$ as obtained from  $H_{\rm CS}(m_0,M,W)$ (see the colorbars in Fig.~\ref{fig:SM_fig2}, \ref{fig:SM_fig3} and \ref{fig:SM_fig4}).

For the non-magnetic case as demonstrated in  Fig.~\ref{fig:SM_fig2}, we find that TPTs between the NI and QSHI phase are found for $(\phi,W)=(0,370)$ and $(\pi,370)$ ($W$ in units of meV). Here the gap between the doubly degenerate bulk valence and conduction bands vanishes while the trivial [topological] gap is observed for  $(\phi,W)=(0,200)$ and $(\pi,200)$ [$(\phi,W)=(0,500)$ and $(\pi,500)$]. In the topological case with $W=500$, we find helical edge modes inside the bulk gap $-25<E<25$ (in  the units of meV) for $\phi=0$ and $\pi$ while there exist no edge mode within the trivial gap for $W=200$. 
Notice that the  critical disorder strength $W_c\approx 360$, separating the QSHI from the NI phases,  (see Fig. 4 (a) of the main text), can be 
approximately traced by the systematic investigations on the band structure  in a stripe geometry with $L_x\to \infty$ considered here.

In the same spirit, for magnetic field $gM=30$ meV, the TPTs between NI and QAHI phase occur at $W_{c,1}=200$ and $W_{c,2}=500$ for the TPTs separating QAHI from QSCI phase (see Fig.~\ref{fig:SM_fig3}). The size of the  trivial  and topological gap, respectively, for the disorder amplitudes  $W=100$, and  $300$, $400$, $600$ are consistent with Fig. 4 (b) in the main text.  The important point to note here is that the QAHI (QSCI) phase hosts one (two) pair(s) of chiral mode(s) due to TRS breaking. 
In the present case, the QAHI (QSCI) phase supports chiral modes coming from the  lower block Hamiltonian (both the lower and upper block Hamiltonian). 
We repeat the same analysis for $gM=52$ meV in Fig.~\ref{fig:SM_fig4} where the TPT between the QAHI and QSCI phase takes place at $W_{c}=580$. The trivial (topological) gap hosting no (edge) modes are depicted for $W=400$ ($W=700$). Two pairs of chiral edge modes can also be seen inside the topological gap for the QSCI phase.

One also notes that $E(k_x) \to -E(-k_x)$ for $\phi\to \phi +\pi$. 
Together with $W \to -W$ for $\phi\to \phi +\pi$, we can obtain an equivalence, $E(k_x,W) = -E(-k_x,-W)$. This gives rise to the symmetric nature of the phase diagrams under phase averaging in the main text for $\pm E_F$. Hence, we only restrict ourselves to positive values of $E_F$ while investigating the phase diagrams in Figs. 2, 3 and 4 of the main text.

\section{Normalized participation ratio}
\begin{figure}[t]
	\centering
	\subfigure{\includegraphics[width=0.48\textwidth]{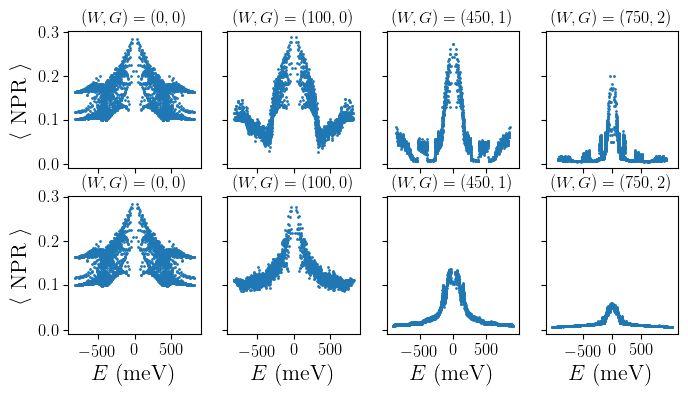}}
	\caption{(Color online) We illustrate the average NPR $I_n$, following Eq.~(\ref{eq:IPR}) for the isolated central system of dimension $30\times 30$, with isotropic and  anisotropic longitudinal quasi-periodicities in upper and lower panel, respectively. The energy window of mobility edge shrinks for the longitudinal case as compared to the isotropic case under substantially strong disorder.    
	We   consider $gM=30$ meV and $(W,G)$ are stated accordingly. }
	\label{fig:SM_IPR}
\end{figure}
We now study the normalized participation ratio (NPR) $I_n$ from the eigenvectors of a system with the spatial dimension $L_x \times L_y$, defined by 
\begin{equation}
    I_n=\Biggl\langle \Bigg(\sum_{i=1}^{pL_x L_y} |\psi_n(i)|^4\Bigg)^{-1} \Biggr\rangle/(pL_x L_y).
    \label{eq:IPR}
\end{equation}
Here the eigenvector for a given energy level $E_n$ is denoted by $\psi_n$. In the present case, we consider the isolated disordered central system $H_{\rm CS}(m_0,M,W)= \sum_{r r'}[ {\mathcal H}_{rr'}(m_0,M) + \epsilon_r \delta_{r, r'}]C^{\dagger}_r C_{r'} $ where ${\mathcal H}_{rr'}(m_0,M)$ is obtained after inverse Fourier transformation of ${ H}(\bm k)$ as given in Eq.~(2) of the main text. The $\langle ... \rangle$ in Eq.~(\ref{eq:IPR}) hence indicates the disorder average and $p=4$ as we have $2$ orbital and $2$ spin degrees of freedom.
For the uniformly delocalized eigenvectors in 2D, $\psi_n$ extends equally over all sites $|\psi_n(i)|^2\sim(p L_x L_y)^{-1}$. The NPR $I_n$ thus approaches unity for the uniformly delocalized state. 
For a localized state with the localization length $\varepsilon$,  one  obtains the    $|\psi_n(i)|^2\sim \varepsilon^{-1}$.  Hence, NPR becomes vanishingly small when
$\varepsilon\ll \sqrt{L_x L_y}$. 
We compute the disorder averaged energy $\langle E_n \rangle$ and study the NPR profile as a function of $\langle E_n \rangle$ depicted in Fig.~\ref{fig:SM_IPR} upper and lower panel for isotropic and longitudinal quasi-periodicity, respectively, with $gM=30$ meV  fixed.
For 1D systems, the NPR turns out to be very important to probe the mobility edge profile, demarcating the localized states from the delocalized states, in the single particle spectrum ~\cite{Biddle11,Modak15,Modak20,Deng19}.

For the 2D case, one can similarly define an energy interval $E_-<E <E_+$ within which the NPR takes higher value for the extended modes.
We start with the clean case $(W,G)=(0,0)$ where a mobility edge is absent (see Fig.~\ref{fig:SM_IPR}). Upon  increases $W$ we observe that
the   delocalized states are symmetrically located around zero energy  (bounded within $E_{\pm}$ i.e., around the centre of the spectrum) and localized states are found to appear away from zero energy (outside the $E_{\pm}$ i.e., around the edge of the spectrum). This refers to the emergence of mobility edge for the disordered 2D system. Note that within our analysis, limited by finite size,  multiple mobility edges might occur (see Fig.~\ref{fig:SM_IPR} $(W,G)=(450,1)$ upper panel). 
The signature of extended modes becomes more pronounced for some intermediate disorder window (see upper and lower panels in Fig.~\ref{fig:SM_IPR} $(W,G)=(100,0)$, $(450,1)$). 
The energy window associated with the mobility edge  is smaller for the anisotropic case compared to the isotropic case for intermediate disorder strength.

Interestingly, we find quantized transport but with $G>2$ stemming from extended bulk modes for the isotropic quasi-periodicity as shown in Fig. 2 of main text. 
This is the same scenario where a mobility edge is promoted.
Therefore, the low energy extended bulk states within the mobility edge might be responsible for the quantized transport for the isotropic case when $E_F$ lies outside of the bulk gap $\Delta$ of the central system. 
For the anisotropic quasi-periodicity in contrast, the mobility edge shrinks
more rapidly with $W$ as compared to the isotropic case.
Therefore, low energy extended bulk modes might not appear  when $E_F$ is outside of the bulk gap $\Delta$.

\section{SCBA based on the continuum model}
In this section, we present the SCBA analysis based on the continuum model, derived from the Eq. (2) in the main text.  
The self-energy $\Sigma$, as formulated by a $2 \times 2$ matrix is given by $(E_F - H -\Sigma)^{-1}=\langle (E_F-{\mathcal  H})^{-1}\rangle$ where 
$\langle \cdots \rangle$ represents the disorder average and  ${H}$ (${\mathcal H}$ ) denotes the $2 \times 2$ ${\bm k}$ space (disordered real space) Hamiltonian.
We expand the Eq. (2) of the main text around the ${\bm \Gamma}=(0,0)$ point to write down the Hamiltonian
\begin{equation}
{ H}(\bm k)=
\begin{pmatrix}
{H}_u(\bm k)
 & 0 \\
0 & {H}_l(\bm k)
\end{pmatrix}, ~{\rm with} ~
{ H}_{l,u}(\bm k)=
\begin{pmatrix}
a_{l,u} & b_{l,u} \\
c_{l,u} & d_{l,u}
\end{pmatrix}
\label{hamiltonian_block} 
\end{equation}
where $a_{l,u}=m_0+Bk^2\mp gM$, $b_{l,u}=v_F (k_y \mp i k_x)$, $c_{l,u}=b^*_{l,u}$ and $d_{l,u}=-a_{l,u}$.  The inverse 
block Hamiltonian $[H_{l,u}(\bm k)]^{-1}$ thus takes the form 
\begin{equation}
[{ H}_{l,u}(\bm k)]^{-1}=\frac{1}{a_{l,u}d_{l,u}-b_{l,u}c_{l,u}}
\begin{pmatrix}
d_{l,u} & -b_{l,u} \\
-c_{l,u} & a_{l,u}
\end{pmatrix},
\label{hamiltonian1_block} 
\end{equation}
The Fourier transformation of the disorder correlation function $C_{m,n} =\left\langle \epsilon_{i,j} \epsilon_{i+m, j+n} \right\rangle=W^2\cos{\bigl[2\pi\eta (m\alpha +n \beta)\bigr]}/8$ is given by 
\begin{eqnarray}
C(\bm k)
&=&\frac{W^2}{8}\sum_{\bm r} e^{i{\bm k}\cdot {\bm r}} \cos{\bigl[2\pi\eta (m\alpha +n \beta)\bigr]} \nonumber \\
&=&\frac{W^2}{16}\bigg[ \delta_{k_x,\alpha Q} \delta_{k_y,\beta Q} + \delta_{k_x,-\alpha Q} \delta_{k_y,-\beta Q} \bigg]    
\end{eqnarray}
with $Q=2\pi \eta$, ${\bm r}=m\alpha \hat{i} +n \beta \hat{j}$. Notice that $m \in [1,L_x]$ and $n \in [1,L_y]$ represent integer numbers.
The self energy, using Eq.~(3) of the main text by setting $\Sigma_{l,u}=0$ in the right hand side, is thus given by \cite{Zimmermann09,Girschik13,Fu21} 
\begin{eqnarray}
    \Sigma_{l,u}&=&\frac{W^2}{16}\Bigg[\frac{1}{M^+_{l,u}} + \frac{1}{M^-_{l,u}}\Bigg]\nonumber \\
    &=& \sum_{k_x=Q^{\pm}_{x},k_y=Q^{\pm}_{y}}  \begin{pmatrix}
    A_{l,u}(k_x,k_y) & B_{l,u}(k_x,k_y) \\
    C_{l,u}(k_x,k_y) & D_{l,u}(k_x,k_y)
\end{pmatrix}.
\end{eqnarray}
We note that the self-energy for the correlated case is thus characteristically different from that of the random disorder where the correlation function $C(\bm k)$ no longer depends on ${\bm k}$. Due to the structure of the correlation function $C(\bm k)$ here, the ${\bm k}$-sum reduces to a $\delta$-function.
Here $M^{\pm}_{l,u}=E_F + i \zeta -H_{l,u}(\pm \alpha Q,\pm \beta Q)$. We denote $Q^{\pm}_x=\pm \alpha Q$ and $Q^{\pm}_y=\pm \beta Q$. One can obtain 
\begin{equation}
    [A,D]_{l,u}=\Big( \frac{W^2}{16} \Big) \frac{1}{{\tilde a}_{l,u}{\tilde d}_{l,u}-{\tilde b}_{l,u}{\tilde c}_{l,u}} [{\tilde d},{\tilde a}]_{l,u}
\end{equation}
where ${\tilde a}$'s are function of $k_x$ and $k_y$ 
with ${\tilde a}_{l,u}=E_F + i \zeta -a_{l,u}$, ${\tilde d}_{l,u}=E_F + i \zeta +a_{l,u}$, ${\tilde b}_{l,u}=-b_{l,u}$ and 
${\tilde c}_{l,u}=-b^*_{l,u}$.  The complete expressions of 
$[A,D]_{l,u}$ are found to be 
\begin{eqnarray}
    A_{l,u}&=&\frac{W^2}{16}\sum_{k_x,k_y} \frac{E_F + i\zeta + (m_0 + Bk^2 \mp gM)}{(E_F + i\zeta)^2 - (m_0 + Bk^2 \mp gM)^2 -v^2_Fk^2} \nonumber \\
    D_{l,u}&=&\frac{W^2}{16}\sum_{k_x,k_y} \frac{E_F + i\zeta -( m_0 + Bk^2 \mp gM)}{(E_F + i\zeta)^2 - (m_0 + Bk^2 \mp gM)^2 -v^2_Fk^2}. \nonumber \\
\end{eqnarray}
Here subscripts $l,u$ in the left hand side correspond to $\mp$ sign in the right hand side.

We are interested in the computation of $\Sigma^{l,u}_0=(A_{l,u}+D_{l,u})/2$ and $\Sigma^{l,u}_z=(A_{l,u}-D_{l,u})/2$ that yield the renormalized mass $\overline {m}^{l,u}_0=m_0+ \delta m_{l,u} $ and chemical potential $\overline {E}^{l,u}_F= E_F+ \delta \mu_{l,u}$ as given by 
\begin{align}
    \delta \mu_{l,u} &=-{\rm Re} [\Sigma^{l,u}_0]\notag\\&=-\Big( \frac{W^2}{8} \Big )\frac{E_F }{E_F ^2-v^2_F x - (m_0 + Bx \mp gM)^2} \notag \\
    \delta m_{l,u}&= {\rm Re} [\Sigma^{l,u}_z]\notag\\&=\Big( \frac{W^2}{8} \Big ) \frac{m_0+Bx \mp gM}{E_F ^2-v^2_F x - (m_0 + Bx \mp gM)^2} \notag 
\end{align}
with $x=Q^2 (\alpha^2 + \beta^2)$.  The TAI phase is supported in the presence of disorder for $\overline {m}^{l,u}_0<0$. The correction in the mass term $\delta m_{l,u}$ turns out to be negative  when $m_0+Bx \mp gM<0$ [$m_0+Bx \mp gM>0$] for $E_F ^2-v^2_F x - (m_0 + Bx \mp gM)^2>0$ [$E_F ^2-v^2_F x - (m_0 + Bx \mp gM)^2<0$]. The analytical findings hint to the situation when both $\delta m_{l}<0$ and $\delta m_{u}<0$  for different combinations of numerator and denominator. Since $\delta m_{l,u}<0$ can be satisfied regardless of the sign of $E_F$ referring to the fact that TAI phases can exist for positive and negative values of $E_F$. 

\section{Topological phase transitions predicted by SCBA based on the lattice model}
We now address the SCBA analysis (Eq. (3) of the main text), based on the lattice Hamiltonian, that accurately complies with the Landauer-B\"uttiker numerical results.  Examining Figs.~2, 3, and 4 of the main text,  
we below discuss the phase boundaries 
following the profiles of $\overline{m}_{0}^{l,u}$ and  $|\overline{E}_F^{l,u}|$. 
The TPTs, separating QAHI from the trivial phase, 
are captured when $\overline{m}_{0}^{l}$ reverses its sign 
simultaneously with $|\overline{E}_F^{l}|=\pm \overline{m}_{0}^{l}$. The QAHI phase is found to be bounded by dashed purple and solid blue in all figures. The QAHI thus appear when $\overline{m}_{0}^{l}<|\overline{E}_F^{l}|<-\overline{m}_{0}^{l}$ and $\overline{m}_{0}^{u}>0$. 
Similarly, the TPTs  between QAHI and QSCI phases, marked by the coincidence of 
solid blue and purple lines,
are associated with  sign changes in $\overline{m}_{0}^{u}$  (while $\overline{m}_{0}^{l}<0$) simultaneously with $|\overline{E}_F^{u}|=\pm \overline{m}_{0}^{u}=0$.
The QSCI/ QSHI phase appears on the right side of the solid purple line. However, the boundary between the QSCI/QSHI and AI phases can not be captured by SCBA. Following the same line of argument, the QSCI/ QSHI phase is expected to be confined within $\overline{m}_{0}^{u,l}<|\overline{E}_F^{u,l}|<-\overline{m}_{0}^{u,l}$ and $\overline{m}_{0}^{u,l}<0$.

\vspace{20pt}


\end{onecolumngrid}

\end{document}